\documentclass[12pt]{article}

\usepackage{amsmath, cite, graphicx}
\usepackage{multirow}
\usepackage{amsfonts}
\usepackage{amssymb}
\usepackage{amscd}
\usepackage{cite}
\usepackage{amsmath}
\usepackage{bbm}

\def\hybrid{\topmargin -20pt    \oddsidemargin 0pt
        \headheight 0pt \headsep 0pt
        \textwidth 6.25in       
        \textheight 9.5in       
        \marginparwidth .875in
        \parskip 5pt plus 1pt   \jot = 1.5ex}

\hybrid

\numberwithin{equation}{section}
\numberwithin{table}{section}

\newcommand{\be}{\begin{equation}}
\newcommand{\ee}{\end{equation}}

\newcommand{\bea}{\begin{eqnarray}}
\newcommand{\eea}{\end{eqnarray}}

\renewcommand{\Re}{\operatorname{Re}}
\renewcommand{\Im}{\operatorname{Im}}

\newcommand\e{\mathrm{e}}
\newcommand\iu{\operatorname{i}}
\newcommand\diff{\mathrm{d}}
\newcommand\vol{\operatorname{vol}}

\newcommand{\ba}{\begin{eqnarray}}
\newcommand{\ea}{\end{eqnarray}}
\newcommand{\ban}{\begin{eqnarray}}
\newcommand{\ean}{\end{eqnarray}}

\begin{document}

\begin{titlepage}

\begin{center}

\rightline{\small CERN-PH-TH/2015-120}

\vskip 2cm

{\Large \bf Consistent truncations of M-theory for general SU(2) structures}

\vskip 1.7cm

{\bf Hagen Triendl} \\

\vskip 1.0cm

{\em Theory Division, Physics Department, CERN, CH-1211 Geneva 23, Switzerland}

\vskip 1.6cm

{\tt hagen.triendl@cern.ch} \\

\end{center}

\vskip 2cm

\begin{center} {\bf ABSTRACT } \end{center}
In seven dimensions any spin manifold admits an $SU(2)$ structure and therefore very general M-theory compactifications have the potential to allow for a reduction to $N=4$ gauged supergravity. We perform this general $SU(2)$ reduction and give the relation of $SU(2)$ torsion classes and fluxes to gaugings in the $N=4$ theory. We furthermore show explicitly that this reduction is a consistent truncation of the eleven-dimensional theory, in other words classical solutions of the reduced theory also solve the eleven-dimensional equations of motion. This reduction generalizes previous M-theory reductions on Tri-Sasakian manifolds and type IIA reductions on Calabi-Yau manifolds of vanishing Euler number. Moreover, it can also be applied to compactifications on certain $G_2$ holonomy manifolds and to more general flux backgrounds.

\vfill

\today

\end{titlepage}

\tableofcontents

\section{Introduction}

It has been argued that the vanishing of certain topological indices of the compactification manifold restricts the appearance of string corrections due to the appearance of additional (spontaneously broken) supercurrent \cite{KashaniPoor:2013en}. This was exemplified in \cite{KashaniPoor:2013en} for Calabi-Yau compactifications of type II for the case of vanishing Euler number.
The vanishing Euler number ensures the existence of an $SU(2)$ structure on the Calabi-Yau manifold. Reduction on such an $SU(2)$ structure  leads to $N=4$ gauged supergravities \cite{ReidEdwards:2008rd,Triendl:2009ap,Louis:2009dq,Danckaert:2011ju,KashaniPoor:2013en,Grimm:2014aha}, which can be conveniently described by the embedding tensor formalism \cite{deWit:2005ub,Schon:2006kz}.
Non-renormalization theorems of $N=4$ supergravity then explain the vanishing of perturbative string corrections for these compactifications and lead to the conjecture that also certain non-perturbative string corrections must vanish for these backgrounds. In the discussed case of Calabi-Yau manifolds with vanishing Euler number it could be shown by applying mirror symmetry that this conjecture indeed is true.

These findings suggest that spontaneously broken supercurrents play a far more important role in string compactifications than considerations of effective actions would suggest. In particular this suggests that there is a general scheme to understand string corrections for general G-structure backgrounds. The most pressing question is whether spontaneously broken supercurrents can also restrict string corrections when only $N=1$ remains unbroken.

A particularly interesting case to address this question are M-theory compactifications to four dimensions. It has been known for a long time that any seven-dimensional spin manifold admits an $SU(2)$ structure \cite{ethomas,Friedrich:1997,Kaste:2003zd}. This suggests that we should be able to find for many such M-theory compactifications a reduction to $N=4$ gauged supergravity, which might give strong constraints on membrane instanton corrections in these backgrounds. In this paper we will perform such an $SU(2)$ structure reduction to four dimensions and determine the corresponding $N=4$ gauged supergravity by identifying the corresponding gaugings.

We show that the reduction performed in this paper is in fact a consistent truncation. Consistent truncations to gauged supergravities have been performed for many particular AdS backgrounds, see for instance \cite{Cassani:2010uw,Gauntlett:2010vu,Cassani:2010na,Bena:2010pr,OColgain:2011ng,Cassani:2011fu,Cassani:2012pj}. Our reduction generalizes the known M-theory reductions to $N=4$ gauged supergravity. Therefore it might help to understand more general four-dimensional AdS backgrounds.

The given reduction is applicable both to compactifications to Minkowski and AdS spacetimes. A particularly interesting application would be to understand the corrections to M-theory compactifications on $G_2$ manifolds: Some of the known Joyce manifolds of \cite{joyce} are dual heterotic Calabi-Yau backgrounds with vanishing Euler number \cite{Papadopoulos:1995da}, where the techniques of \cite{KashaniPoor:2013en} could be used.

The paper is organized as follows. In Section~\ref{sec:SU2in7} we discuss general $SU(2)$ structure manifolds in seven dimensions and thereby set the stage for performing the reduction. In Section~\ref{sec:reduction} we will make the reduction ansatz and then perform the $SU(2)$ reduction to four dimensions. The embedding tensor components of the corresponding $N=4$ gauged supergravity are identified in Section~\ref{sec:matching}. In Section~\ref{sec:consistent} we discuss the consistency of the truncation, and in Section~\ref{sec:vacua} we make contact with some classes of AdS vacua in the literature. Some of the technical details as well as our conventions regarding $N=4$ gauged supergravity are presented in three appendices.

\section{$SU(2)$ structures on seven-manifolds}
\label{sec:SU2in7}
Let us start by introducing the concept of an $SU(2)$ structure on a seven-dimensional manifold $Y$. On a seven-dimensional manifold the spinor bundle is eight-dimensional. We will be interested in the splitting $SO(7) \to SO(3) \times SO(4)$ which reads for the corresponding spin groups ${\rm Spin}(7) \to SU(2) \times SU(2) \times SU(2)$.
A seven-dimensional $SU(2)$ structure manifold $Y$ admits four nowhere vanishing spinors $\eta_{i \hat \imath}$, $i,\hat \imath=1,2$, whose norm we fix by imposing
\begin{equation}\label{eq:spinor_norm}
\bar \eta^{j\hat \jmath} \eta_{i\hat \imath} = \delta^j_i \delta^{\hat \jmath}_{\hat \imath} \ .
\end{equation}
These four spinors arise under the breaking ${\rm Spin}(7) \to SU(2)_3 \times SU(2)_4 \times SU(2) \to SU(2)$ by
\begin{equation}
 ({\bf 8}) \to ({\bf 2},{\bf 2},{\bf 1}) \oplus ({\bf 2},{\bf 1},{\bf 2}) \to 4 ({\bf 1}) \oplus 2 ({\bf 2}) \ .
\end{equation}
We denoted in \eqref{eq:spinor_norm} the index of the broken $SU(2)_3$ that is related to the $SO(3)$ by $i$ and the index of the broken $SU(2)_4$ inside $SO(4)$ by $\hat \imath$. The third $SU(2)$ subgroup is the (unbroken) structure group.

Based on these spinors we can introduce a $SU(2)$ triple of (real) two-forms $J^{\hat a}$, ${\hat a}=1,2,3$, and a triple of (real) one-forms $K^a$, $a=1,2,3$, via
\begin{equation} \label{eq:def_JaK}
 J^{\hat a} =  \iu \sqrt{\tfrac{3}{2}} (\sigma^{\hat a})^{\hat \imath}{}_{\hat \jmath} \bar \eta^{i\hat \jmath} \gamma_{mn} \eta_{i\hat \imath} \diff x^m \wedge \diff x^n \ , \qquad K^a = (\sigma^a)^i{}_j \bar \eta^{j\hat \imath} \gamma_m \eta_{i\hat \imath} \diff x^m \ .
\end{equation}
The Fierz identities for these spinors can now  be parametrized by
\begin{equation} \label{spinor_bilinears} \begin{aligned}
\eta_{i\hat \imath} \otimes \bar \eta^{j\hat \jmath} \, = \, &  \tfrac18 \left( (1+ \gamma_{(3)}) \delta^j_i + \iu (\sigma^a)^i{}_j (K^a_m \gamma^{m} - \epsilon^{abc} K^b_m K^c_n \gamma^{mn}) \right) \\ & \cdot \left(  (1 - \gamma_{(4)})\delta_{\hat \imath}^{\hat \jmath} +  \sqrt{\tfrac{2}{3}}\iu(\sigma_{\hat a})_{\hat \imath}^{\hat \jmath} J^{\hat a}_{pq} \gamma^{pq} \right) \  .
\end{aligned}\end{equation}
Taking the products of these bilinears and using \eqref{eq:spinor_norm} yields the relations
\begin{equation}\label{eq:JaJb}
\tfrac12 J^a \wedge J^b = \delta^{ab} \vol_4
\end{equation}
and
\begin{equation}\label{eq:K_compatible}
\epsilon^{abc} K^a \wedge K^b \wedge K^c  ={\rm vol}_3 \ .
\end{equation}
The Fierz identities guarantee also the existence of an almost product structure $P: TY \rightarrow TY$, $P^2=id$, on the manifold, defined locally via
\begin{equation}
\label{eq:almost_product_structure}
P_m{}^n = 2  K^a_m \hat K_a^{n}  - \delta_m^{n} \ .
\end{equation}
where the vectors $\hat K_a$ are defined by
\begin{equation}
K^{a}(\hat K_b) = \delta^a_b \ , \qquad J^{\hat a}(\hat K^b, \cdot) =0 \ .
\end{equation}
The eigenspaces $T_3 Y$ and $T_4 Y$ of $P$ to the eigenvalues $+1$ and $-1$ respectively yield a global decomposition of the tangent space,
\begin{equation}\label{eq:tangent_space_splitting}
T Y = T_3 Y \oplus T_4 Y \ .
\end{equation}
The subbundle $T_3 Y$ is trivial, spanned by $\hat K_a$. By definition of $P$, the $K^a$ ($J^{\hat a}$) are trivial on $T_4 Y$ ($T_3 Y$).
Note that the splitting \eqref{eq:tangent_space_splitting} and the definition of such an $SU(2)_3$ triple of one-form $K^a$ and an $SU(2)_4$ triple of two-forms $J^{\hat a}$, ${\hat a}=1,2,3$, which satisfy the conditions \eqref{eq:JaJb} and \eqref{eq:K_compatible}, allow for a definition of the $SU(2)$-structure with no reference to spinors.

Let us now discuss the frame bundle over spacetime times $Y$. We choose a section (vielbein)\footnote{Eq.\ \eqref{eq:K_compatible} implies that the $K^a$ can be chosen as components of the vielbein.}
\begin{equation}\label{vielbein}
 e^A = (e^\mu, K^a, e^\alpha) \ ,
\end{equation}
where the $e^\mu$ live in spacetime and depend only on spacetime coordinates. In contrast, the $K^a = k^a_b (v^b + G^a)$ consist of one-forms $v^a$ in $T^*_3$ and spacetime gauge fields $G^a$ (the Kaluza-Klein vectors) that parameterize the fibration of $T^*_3$ over spacetime, as well as the coefficient $k^a_b$, which is a spacetime scalar.\footnote{Due to the mixed spacetime/internal components of the ten-dimensional metric, the components $K^a$ of the vielbein are not purely internal. We will nevertheless retain the same nomenclature as in (\ref{eq:def_JaK}) for simplicity.} Furthermore, the $e^\alpha$ are one-forms on $T^*_4$ such that
\begin{equation}\label{eq:Jexpl}
 J^{\hat a} = \tfrac12 (I^{\hat a})^{\alpha}_{\beta} e^\alpha \wedge e^\beta \ ,
\end{equation}
with constant coefficients $(I^{\hat a})^{\alpha}_{\beta}$ that are the generators of the $SU(2)_4$ algebra of complex structures on the frame bundle, i.e.\
\begin{equation} \label{eq:IaIb}
 (I^{\hat a})^{\alpha}_{\gamma} (I^{\hat b})^{\gamma}_{\beta} = \epsilon^{{\hat a}{\hat b}{\hat c}} (I^{\hat c})^{\alpha}_{\beta} - \delta^{{\hat a}{\hat b}} \delta^{\alpha}_{\beta}  \ .
\end{equation}
Similarly, the $\tilde I^a$ are generators of $SO(3)$ given by $(\tilde I^a)^b_c = \epsilon^{abc}$.
The dual vielbein to \eqref{vielbein} is
\begin{equation}
 \hat e_A = (\hat e_\mu, \hat K_a, \hat e_\alpha) = (\partial_\mu- G_\mu^a \hat v_a, (k^{-1})^b_a \hat v_b, \hat e_\alpha)\ ,
\end{equation}
where $\hat v_a$ are the vector fields dual in $T^*_3$ to the vielbein component $v^a$.

Next, we consider the Levi-Civita connection one-form $\Omega$, which is the unique torsion-free connection satisfying the Maurer-Cartan equation
\begin{equation}\label{eq:MaurerCartan}
D e = \diff e + \Omega \wedge e = 0 \ .
\end{equation}
The corresponding curvature two-form is defined by
\begin{equation} \label{eq:curv}
R = \diff \Omega + \Omega \wedge\Omega \ .
\end{equation}
The Ricci tensor (in flat indices) is defined by contraction with the dual vielbein,
\begin{equation} \label{eq:Ricci}
{\rm Ric}_{AB} = R^C_A (\hat e_C,\hat e_B) \ ,
\end{equation}
and the Ricci scalar as its trace
\begin{equation} \label{eq:rscalar}
r_{11} = {\rm Ric}_{AB}\delta^{AB} \ .
\end{equation}
Let us decompose the eleven-dimensional connection under the breaking of the Lorentz group $SO(1,10) \to SO(1,3) \times SO(3)\times SO(4)$ as
\begin{equation}\begin{aligned}
 {\bf 55} &&= && ({\bf 6},{\bf 1},{\bf 1}) && \oplus &\ ({\bf 4},{\bf 1},{\bf 4}) && \oplus &\ ({\bf 4},{\bf 3},{\bf 1}) && \oplus &\ (({\bf 1},{\bf 3},{\bf 1}) \oplus ({\bf 1},{\bf 3},{\bf 4}) \oplus ({\bf 1},{\bf 1},{\bf 6}))\ ,\\
 \Omega &&= && \omega &&+&\quad [\lambda] &&+&\quad [\gamma] &&+&\quad \Theta \ ,
\end{aligned}\end{equation}
where we have called the full $SO(7)$ connection $\Theta$. Using $SO(3) \equiv (SU(2)_3)/\mathbb{Z}_2$ and $SO(4) \equiv (SU(2)_4 \times SU(2))/\mathbb{Z}_2$, we can further decompose the adjoint representation of $SO(7)$ and thus the internal connection $\Theta$ as
\begin{equation}\label{so6decomp_connection}\begin{aligned}
 so(7) &&= && su(2)_3&& \oplus &\ su(2)& \oplus & \ su(2)_4 & \oplus &\ ({\bf 3},{\bf 2},{\bf 2}) \ ,\\
 \Theta &&= && [\phi^a] &&+&\quad  \theta &+& \quad [\psi^{\hat a}] &+&\quad  [\tau] \ ,
\end{aligned}\end{equation}
where $su(2)$ is the adjoint of the $SU(2)$ structure group, $su(2)_4$ is spanned by the $I^{\hat a}$ and $su(2)_3$ by $(I^a)^b_c = \epsilon^{abc}$.
Using this decomposition and the vielbein \eqref{vielbein}, the Maurer-Cartan equations \eqref{eq:MaurerCartan} read in components
\begin{equation}\label{eq:MCexplicit}\begin{aligned}
 \diff e^\mu + \omega^\mu_\nu \wedge e^\nu + \lambda^\mu_\alpha \wedge e^\alpha + \gamma^\mu_a \wedge K^a & =  0 \ , \\
 \diff K^a + \epsilon^{abc} \phi^c \wedge K^b + \tau^a_\alpha \wedge e^\alpha + \gamma_\mu^a \wedge e^\mu & =  0 \ , \\
 \diff e^\alpha + \theta^\alpha_\beta \wedge e^\beta + (I^{\hat a})^\alpha_\beta \psi^{\hat a} \wedge e^\beta + \tau^\alpha_a \wedge K^a + \lambda_\mu^\alpha \wedge e^\mu & =  0 \ .
\end{aligned}\end{equation}

Note that the connection component $\theta$ is the torsionful $SU(2)$ connection. Its internal torsion two-form $T$ can be expressed in terms of the other components of $\Theta$. On $T_3 Y$ the internal torsion is given by $T^a = \diff K^a$ and the component on $T_4$ is
\begin{equation} \label{torsion4d}
 T^\alpha = \diff e^\alpha + \theta^\alpha_\beta \wedge e^\beta =  - (I^{\hat a})^\alpha_\beta \phi^{\hat a} \wedge e^\beta - \tau^\alpha_a \wedge K^a \ .
\end{equation}

Similar to the connection one-form we can also decompose the Ricci tensor group-theoretically. In particular, we are interested in the `symmetric' representation $S^2 T^* Y$, which decomposes as
\begin{equation}\label{symmetric_decomp}\begin{aligned}
 S^2 T^* Y \, = \, &  S^2_0 T^*_3 Y \oplus \mathbb{R} g_3^{(0)}\oplus S^2_0 T^*_4 Y \oplus \mathbb{R} g_4^{(0)} \oplus (T^*_3 Y \otimes T^*_4 Y) \\
 = \, &  ({\bf 5},{\bf 1},{\bf 1}) \oplus ({\bf 1},{\bf 1},{\bf 1}) \oplus ({\bf 1},{\bf 3},{\bf 3})\oplus ({\bf 1},{\bf 1},{\bf 1})\oplus ({\bf 3},{\bf 2},{\bf 2}) \ .
\end{aligned}\end{equation}
Here, the $({\bf 1},{\bf 3},{\bf 3})$ representation $S^2_0 T^*_4$ is spanned by the products of generators of $su(2)_4$ and $su(2)$. In other words, since the elements of $su(2)$ and $su(2)_4$ commute, the representation can be written as
\begin{equation}\label{S2T4representation}
 S^2_0 T^*_4 = \{ (I^{\hat a})^\alpha_\gamma ({\cal I}_{\hat a})^\gamma_\beta | {\cal I}_{\hat a} \in su(2), {\hat a}=1,2,3 \} \ .
\end{equation}

\section{Dimensional reduction from M-theory}
\label{sec:reduction}

In this section, we will reduce the eleven-dimensional supergravity action
\begin{equation} \label{eq:actionM} \begin{aligned}
S_{11} = & \ \tfrac{1}{2\kappa_{11}^2}  \int_{11} ((\ast_{11} 1) r_{11}  -\frac12 G_4 \wedge \ast_{11} G_4) \\ & -\tfrac16 \int_{11} G_4 \wedge G_4 \wedge C_3 \ ,
\end{aligned} \end{equation}
to four dimensions. Here, $G_4 = \diff C_3$ is the form field strength of the three-form gauge field.

\subsection{The reduction ansatz}
\label{sec:cons_trunc}

The almost product structure \eqref{eq:almost_product_structure} on $Y$ will play a central role in the choice of our reduction ansatz. $T_3$ has trivial structure group and is therefore parallelizable. We hence introduce a basis of three global one-forms $v^a$, $a=1,2,3$, on this subbundle, yielding three one-forms, three two- and a three-form (their wedge products) as expansion forms. On $T_4$  our ansatz similarly to \cite{KashaniPoor:2013en} contains $SU(2)$ singlets and triplets. It is easily checked that $SU(2)$ doublets exactly correspond to odd forms on $T_4$. Therefore, the ansatz will consist of two-forms $\omega^I, I=1,\dots, n,$ that all square to the same volume form $\vol^{(0)}_4$ on $T_4$, i.e.\
\begin{equation} \label{eq:omegavol4}
 \tfrac12 \omega^I \wedge \omega^J = \eta^{IJ} \vol^{(0)}_4 \ ,
\end{equation}
where $\eta$ is a metric with signature $(3,n-3)$, reflecting the number of singlet and triplet representations as discussed above. Furthermore, we include all wedge products of $\omega^I$ and $v^a$ in the reduction ansatz. For instance, we expand the forms $J^{\hat a}$ and $K^a$ of \eqref{eq:def_JaK} that specify the $SU(2)$ structure in the set of modes $\omega^I, I=1,\dots, n,$ and $v^a$, $a=1,2,3$, i.e.\
\begin{equation} \label{eq:param_forms}
 J^{\hat a} = \e^{\rho_4/2} \zeta^{\hat a}_I \omega^I \ , \qquad K^a = \e^{\rho_3/3} k^a_b (v^b + G^b) \ ,
\end{equation}
where $\operatorname{det}(k)= 1$. Furthermore, we fix $\epsilon^{abc} v^a \wedge v^b \wedge v^c  = \vol^{(0)}_3$.
Note that a consequence of the second equation in \eqref{eq:param_forms} is that
\begin{equation} \label{eq:k_det}
 k^a_d k^b_e k^c_f \epsilon^{def} = \epsilon^{abc} \ , \qquad  k^a_d k^b_e k^c_f \epsilon_{abc}  = \epsilon_{def} \ ,
\end{equation}
which in particular means that $\det(k) = 6$.


Note that the presence of internal one-forms in our ansatz gives rise to Kaluza-Klein vectors $G^i$, i.e.\ mixed spacetime and internal components of the ten-dimensional metric. The expansion coefficients $\zeta^{\hat a}_I$, $\rho_4$, $\rho_3$ and $k^a_b$ depend on the spacetime coordinates and give rise to scalar fields in four dimensions.
Furthermore, \eqref{eq:JaJb} yields the relations
\begin{equation}\label{eq:zeta_ab}
\zeta^{\hat a}_I \eta^{IJ} \zeta^{\hat b}_J = \delta^{{\hat a}{\hat b}} \ ,
\end{equation}
and
\begin{equation}
\vol_4 = \e^{\rho_4} \vol^{(0)}_4 \ , \qquad \vol_3 = \e^{\rho_3} \vol^{(0)}_3 \ .
\end{equation}
The four-dimensional fields $\rho_{3/4}$ describe the volume moduli of $T_{3/4}$ while the $\zeta^{\hat a}_I$ describe the $SU(2)$-structure geometry and $k^a_b$ describes the three-dimensional geometry.

We can also expand the three-form gauge field in terms of this basis. This gives
\begin{equation}\label{eq:form_field} \begin{aligned}
C_3 = & \hat C + \hat C_a \wedge (v^a + G^a) + C_I \wedge \omega^I + \tfrac12 \epsilon_{abc} C^a \wedge (v^b +G^b) \wedge (v^c + G^c)  \\ & + \tfrac16  c_{0} \epsilon_{abc} (v^a + G^a) \wedge (v^b +G^b) \wedge (v^c + G^c) +c_{aI} (v^a + G^a) \wedge \omega^I \ .
\end{aligned}\end{equation}
We will also describe fluxes in this setup, therefore our ansatz for the four-form field strength will be
\begin{equation}\label{eq:formfieldstrength}
 G_4 = G^{\rm flux}_4 + \diff C_3 \ ,
\end{equation}
where we define
\begin{equation}\label{eq:flux}
 G^{\rm flux}_4 = \e^{-3 \phi- \rho_3} f_0 \vol_4({\rm Mink}) + g_0 \vol^{(0)}_4 + \tfrac12 g^a_{I} \epsilon_{abc} v^b  \wedge v^c \wedge \omega^I \ ,
\end{equation}
with $f_0$, $g_0$ and $g^a_{I}$ being constants,
and we demand the Bianchi identity
\begin{equation}\label{eq:Bianchi}
\diff G_4 = 0 \ .
\end{equation}
Furthermore, $G^{\rm flux}_4$ is only defined up to an exact piece, so that only a subset of the numbers $(g_0, g_{aI})$ are actual flux numbers. Also, note that the flux piece in \eqref{eq:flux} proportional to $\vol_4({\rm Mink})$ has a dependence on the volume factors because it originates from dual seven-form flux
\begin{equation}
\label{eq:G7flux}
G^{\rm flux}_7 = f_0 \vol^{(0)}_3 \wedge \vol^{(0)}_4  \ .
\end{equation}
Note that the flux piece in \eqref{eq:flux} proportional to $\vol_4({\rm Mink})$ can be absorbed in $\diff \hat C$ but will reoccur later when we introduce dual fields. We discuss this seven-form flux again in Appendix \ref{app:dual}.

The $J^{\hat a}$ and the $K^a$ in general define the Hodge star, which splits into a spacetime component and two components $\ast_3$ and $\ast_4$ acting on forms on $T_3$ and $T_4$, respectively. The precise form of $\ast_3$ and $\ast_4$ is fixed by
\begin{equation}
 \ast_3 K^a = \tfrac12 \epsilon_{abc} K^b \wedge K^c \ , \qquad  \ast_3 1 = {\rm vol}_3 \ , \qquad \ast_4 J^{\hat a} = J^{\hat a} \ .
\end{equation}
The latter can be translated into
\begin{equation}
 \ast_4 \omega^I = (2 \zeta^{a\, I} \zeta^a_J - \delta^I_J) \omega^J = H^{I}_J \omega^J \ , \qquad \ast_4 1 = \vol_4 = \e^{\rho_4} \vol_4^{(0)} \ .
\end{equation}
In the following reduction, we will assume that the internal volume is normalized,
\begin{equation}
\int_7 \tfrac16 \epsilon^{abc} v^a \wedge v^b \wedge v^c \wedge \vol_4^{(0)} = \frac{\kappa_{11}^2}{\kappa_4^2} \ .
\end{equation}

To perform the reduction, we must next specify the differentials of the expansion forms $\{ v^i, \omega^I \}$. As remarked above, we will require that the differential algebra of modes they span closes, i.e.\
\begin{equation} \label{eq:truncation_ansatz}\begin{aligned}
 \diff v^a = & \tfrac12 t^{ab} \epsilon_{bcd} v^c \wedge v^d + t^a_I \omega^I \ , \\
 \diff \omega^I = & \tilde T^I_{aJ} v^a \wedge \omega^J \ .
\end{aligned}\end{equation}
Here, the coefficients $t^{ab}$, $t^a_I$ and $\tilde T^I_{aJ}$ are constants that parameterize the $SU(2)$ structure reduction ansatz for a particular manifold.
In particular we exclude any terms on the right-hand side of the above equations involving $SU(2)$ doublets.
Note also that in the second equation of \eqref{eq:truncation_ansatz} a possible term proportional to $v^1 \wedge v^2 \wedge v^3$ is immediately set to zero by the constraint that $\diff (\omega^I \wedge \omega^J \wedge \omega^K)=0$.

The $t^{ab}$, $t^a_I$ and $\tilde T^I_{aJ}$ specify the torsion classes of $Y$. We choose them and hence the torsion classes of $Y$ to be constant. These constants are constrained by the fact that the exterior derivative squares to zero and the integral of $\diff (v^a \wedge v^b\wedge \omega^I \wedge \omega^J)$ over Y should vanish. The constraints are encapsulated by algebraic relations, given by
\begin{equation}\label{eq:quadratic_constraints} \begin{aligned}
t^{ab} \epsilon_{bcd} t^{cd} & = 0 \ , \\
 \epsilon_{bcd} t^{ac} t^d_I  + t^a_J \tilde T^J_{bI} & = 0 \ , \\
 \tilde T^I_{aJ} \eta^{JK} t^a_K & = 0 \ , \\
  \tilde T^I_{bJ} t^{ba}  - \epsilon^{abc} \tilde T^I_{bK} \tilde T^K_{cJ} & = 0 \ , \\
\epsilon_{abc} t^{bc} \eta^{IJ} + \tilde T^I_{aK}\eta^{KJ} + \tilde T^J_{aK}\eta^{KI} & = 0 \ .
\end{aligned}\end{equation}
The last equation determines the symmetric part of $\tilde T^I_{jK}\eta^{KJ}$, $j=1,2$, so that
\begin{equation}\label{eq:decomp_T}
 \tilde T^I_{aK}\eta^{KJ} = T^I_{aK}\eta^{KJ}- \tfrac12 \epsilon_{abc} t^{bc} \eta^{IJ} \ ,
\end{equation}
where $T^I_{aK}$ is a triple of $so(3,n-3)$ matrices, i.e.\
\begin{equation}
T^I_{aK} \eta^{KJ} + T^J_{aK}\eta^{KI} = 0 \ .
\end{equation}
The fourth condition just states that the $T^I_{aJ}$ form an $so(3,n-3)$ subalgebra $S$ defined by
\begin{equation}
  \epsilon^{abc} T^I_{bK} T^K_{cJ} =   t^{ba} T^I_{bJ}  \ .
\end{equation}
In particular, $S$ is Abelian if $t^{ab}=0$ and simple if $t^{ab}$ has rank three.
The remaining condition is
\begin{equation}\label{eq:quadrem}
   T^J_{aI} t^b_J = \tfrac12 \epsilon_{acd}(t^{cd} t^b_I - 2 t^{bc} t^d_I )\ .
\end{equation}
If $t^{ab}$ is zero, the $t^a_I$ are invariant under $S$. If $t^{ab}$ is non-zero, the $t^a_I$ form a non-trivial representation under $S$.
The Bianchi identity \eqref{eq:Bianchi} also leads to constraints on the flux numbers appearing in \eqref{eq:flux}, given by
\begin{equation}\label{eq:quadflux} \begin{aligned}
 \epsilon_{abc} (\eta^{IJ} g^{b}_{I} t^{c}_J - g_0 t^{bc}) & = 0 \ , \\
\tfrac12 \epsilon_{abc} g^a_{I} t^{bc} + g^a_{J} T^J_{aI} & = 0 \ .
\end{aligned} \end{equation}

\subsection{Reduction of gravity}
In this section we dimensionally reduce the gravitational term in the eleven-dimensional supergravity action \eqref{eq:actionM},
\begin{equation} \label{eq:EHaction}
S_{\rm grav} = \tfrac{1}{2\kappa_{11}^2}  \int_{11}  (\ast_{11} 1) r_{11} \ .
\end{equation}
For this we have to compute the eleven-dimensional Ricci scalar in terms of the ansatz \eqref{eq:param_forms}. We start by computing the connection $\Omega$ from the Maurer-Cartan equations \eqref{eq:MCexplicit} and \eqref{eq:truncation_ansatz} in Appendix~\ref{sec:connection}.
There we find for the components of the connection
\begin{equation}\label{eq:con_exp} \begin{aligned}
\tau^a_\alpha = & \tfrac12 \e^{\rho_3/3} k^a_b t^b_I \omega^I_{\alpha\beta} e^\beta  - \tfrac14 \e^{-\rho_3/3} (k^{-1})^b_a \epsilon_{bcd} t^{cd} e^\alpha  + \tfrac12 \e^{-\rho_3/3} \e^{\rho_4/2} (k^{-1})^b_a \zeta^{\hat a}_I T^I_{bJ}  P^J_K \omega^K_{\alpha \gamma} ( I^{\hat a})^\gamma_\beta e^\beta \ ,\\
\gamma^a_\mu = & \e^{\rho_3/3} k^a_b D_{[\mu} G^b_{\nu]} e^\nu + \tfrac12 ((k^{-1})_b^c  D_\mu k^{a}_c +(k^{-1})_a^c D_\mu k^{b}_c ) K^b +\tfrac13 (D_\mu \rho_3) K^a \ , \\
\phi^a = & - \e^{-\rho_3/3} (k^a_c k^b_d t^{cd} - \tfrac12 \delta^{ab} k^e_c k^e_d t^{cd}) K^b - \tfrac12 \epsilon^{abc} (k^{-1})_b^d D k_d^c \ ,\\
\Psi^{\hat a} = & \tfrac14 \e^{-\rho_3/3} \epsilon^{\hat a \hat b \hat c} \zeta^{\hat b}_I \zeta^{\hat c \, J}  T^I_{bJ} (k^{-1})^b_a K^a - \tfrac14 \e^{\rho_3/3} \e^{-\rho_4/2} \zeta^{\hat a\, I} t^b_I k^a_b K^a - \tfrac14 \epsilon^{\hat a\hat b \hat c} \zeta^{\hat b}_I \zeta^{\hat c\, J} T^I_{bJ} G^b \ , \\
\lambda^\alpha_\mu = & \tfrac14 D_\mu \rho_4 e^\alpha - \tfrac12 \e^{\rho_4/2} D_\mu \zeta^{\hat a}_I \omega^I_{\alpha \gamma} (I^{\hat a})^\gamma_\beta e^\beta   \ , \\
\omega^\mu_\nu = & \hat \omega^\mu_\nu - \e^{\rho_3/3} D_{[\mu} G^b_{\nu]} k^a_b K^a\ ,
\end{aligned}\end{equation}
with the projector $P^I_J = \delta^I_J - \zeta^{\hat a\, I} \zeta^{\hat a}_J$ and the covariant derivatives given by
\begin{equation}\label{eq:cov_metric}\begin{aligned}
 D G^a = & \diff G^a + \tfrac12 t^{ab} \epsilon_{bcd} G^c \wedge G^d \ , \\
 D\rho_3 = & \diff \rho_3 - \epsilon_{abc} t^{bc} G^a\ , \\
 D \rho_4 = & \diff \rho_4 + \epsilon_{abc} t^{bc} G^a  \ , \\
 D k^a_b = & \diff k^a_b - (k^a_{c} \epsilon_{bde} - \tfrac13 k^a_{b} \epsilon_{cde})  t^{cd} G^e \ , \\
 D \zeta^{\hat a}_I = & \diff \zeta^{\hat a}_I - \zeta^{\hat a}_K T^K_{bJ} P^J_I G^b \ .
\end{aligned}\end{equation}
In Appendix \ref{sec:Ricci} we compute from this connection the components of the ten-dimensional Ricci curvature. For the reduction we only need the Ricci scalar $r_{11}$, given by
\begin{equation}\begin{aligned}
r_{11} = & r_4 - 2 \nabla^\mu (D_\mu \rho_4) - 2 \nabla^\mu (D_\mu \rho_3)
\\ &
 - \tfrac54 (D_\mu \rho_4) (D^\mu \rho_4) - \tfrac43 (D_\mu \rho_3)(D^\mu \rho_3) - 2 D^\mu \rho_4 D_\mu \rho_3
  \\ &
 -\tfrac12  {\rm tr} [k^{-1} \cdot (D_\mu k) \cdot k^{-1} \cdot (D^\mu k)]
 - \tfrac12 {\rm tr} [(D_\mu k)^T \cdot (g_3^{-1}) \cdot (D^\mu k)] \\ &
 +(D_\mu \zeta^{\hat a}_I) (D^\mu \zeta^{\hat a \, I})
  - \e^{2\rho_3/3} g_{3 \, ab}  D_{[\mu} G^a_{\nu]} D^{\mu} G^{b \,\nu}  \\ &
 + \tfrac32 \e^{-2 \rho_3/3} (g_{3\, ab}t^{ab})^2
  - \e^{-2 \rho_3/3} g_{3\, ab} g_{3\, cd} t^{(ac)} t^{(bd)}    - \tfrac{3}{4} \e^{-2 \rho_3/3}  g^{ab}_{3} \epsilon_{acd} \epsilon_{bef} t^{cd} t^{ef} \\ &
 - \e^{2\rho_3/3} \e^{-\rho_4} g_{3\, ab} t^a_I H^{IJ} t^b_J
 - \tfrac13 \e^{- 2\rho_3/3} g_3^{ab} \zeta^{\hat a}_I \zeta^{\hat a}_J T^I_{aK} P^{KL} T^J_{bL}  \\ &
- 2 \e^{-\rho_4/2} \epsilon^{\hat a \hat b \hat c} \zeta^{\hat a \, I} \zeta^{\hat b}_J \zeta^{\hat c \, K} t^a_I  T^J_{aK}  \ ,
\end{aligned}\end{equation}
where $r_4$ is the four-dimensional Ricci scalar. Note that we can rewrite
\begin{equation}\begin{aligned}
 &-\tfrac12  {\rm tr} [k^{-1} \cdot (D_\mu k) \cdot k^{-1} \cdot (D^\mu k)]
 - \tfrac12 {\rm tr} [(D_\mu k)^T \cdot (g_3^{-1}) \cdot (D^\mu k)] \\
 = & - \tfrac14 {\rm tr} [(D_\mu g_3) \cdot (g_3^{-1}) \cdot (D^\mu g_3) \cdot g_3^{-1}]  \ ,
\end{aligned}\end{equation}
with the definition of the covariant derivative as
\begin{equation}
 D g_{3\, ab} =  \diff g_{3\, ab} - (g_{3\, ac} \epsilon_{bde} + g_{3\, cb} \epsilon_{ade} - \tfrac23 g_{3\, ab} \epsilon_{cde})  t^{cd} G^e \ ,
\end{equation}

The eleven-dimensional volume form includes a prefactor $\e^{\rho_4 + \rho_3}$ that describes the scaling of the internal volume. Thus the reduction of the eleven-dimensional Einstein-Hilbert action to four dimensions in \eqref{eq:EHaction} is
\begin{equation}
S_{\rm grav} = \tfrac{1}{2\kappa_{4}^2}  \int_{4}  (\ast_{4} 1) \e^{\rho_4 + \rho_3} r_{11} \ .
\end{equation}
We perform a Weyl rescaling
\begin{equation} \label{eq:Weyl}
e^\mu \to \e^{- (\rho_3 +\rho_4)/2} e^\mu \ ,
\end{equation}
to bring the four-dimensional Einstein-Hilbert term into its canonical form and get
\begin{equation} \label{eq:actiongrav}\begin{aligned}
S_{\rm grav} = & \tfrac{1}{2\kappa_{4}^2}  \int_{4}  (\ast_{4} 1) \Big( r_4
 - \tfrac{1}{2} (D_\mu \rho_3) (D^\mu \rho_3)
 - \tfrac{3}{4} (D_\mu \phi)(D^\mu \phi) \\ &
- \tfrac14 {\rm tr} [(D_\mu g_3) \cdot (g_3^{-1}) \cdot (D^\mu g_3) \cdot g_3^{-1}] \\ &
 - H^{IJ} (D_\mu \zeta^{\hat a}_I) (D^\mu \zeta^{\hat a}_J)
  - \e^{\phi + \rho_3} g_{3 \, ab} D_{[\mu} G^a_{\nu]} D^{\mu} G^{b \,\nu}  \\ &
 + \tfrac32 \e^{-\phi - \rho_3} (g_{3\, ab}t^{ab})^2
  - \e^{-\phi - \rho_3} g_{3\, ab} g_{3\, cd} t^{(ac)} t^{(bd)}   \\ &
  - \tfrac{3}{4} \e^{-\phi - \rho_3}  g^{ab}_{3} \epsilon_{acd} \epsilon_{bef} t^{cd} t^{ef}
 - \e^{-2\phi + \rho_3} g_{3\, ab} t^a_I H^{IJ} t^b_J\\ &
 - \tfrac13 \e^{-\phi - \rho_3} g_3^{ab} \zeta^{\hat a}_I \zeta^{\hat a}_J T^I_{aK} P^{KL} T^J_{bL}
- 2 \e^{-\tfrac32 \phi } \epsilon^{\hat a \hat b \hat c} \zeta^{\hat a \, I} \zeta^{\hat b}_J \zeta^{\hat c \, K} t^a_I  T^J_{aK} \Big) \ ,
\end{aligned}\end{equation}
where we defined $\phi = \rho_4 + \tfrac23 \rho_3$ with
\begin{equation}
 D \phi = \diff \phi + \tfrac13 \epsilon_{abc} t^{bc} G^a \ .
\end{equation}

\subsection{Reduction of the four-form field strength}
\label{sec:fourformred}
Next we want to reduce the four-form field strength action
\begin{equation} \label{eq:actionG}
S_{G_4} =  \ -\tfrac{1}{4\kappa_{11}^2}  \int_{11} G_4 \wedge \ast_{11} G_4  -\tfrac{1}{12\kappa_{11}^2} \int_{11} G_4 \wedge G_4 \wedge C_3 \ .
\end{equation}
For this we compute the four-form field strength $G_4$, defined in \eqref{eq:formfieldstrength}, using \eqref{eq:truncation_ansatz} and \eqref{eq:flux}. We find
\begin{equation}\label{eq:G4} \begin{aligned}
 G_4 = & (\e^{-3 \phi- \rho_3} f_0 (\ast_4 1)+ \diff \hat C + \hat C_a \wedge DG^a ) + (D \hat C_a - \epsilon_{abc} C^b \wedge DG^c)  \wedge (v^a + G^a) \\ &
 + (DC_I + c_{aI} DG^a) \wedge \omega^I + \tfrac12 \epsilon_{abc} (D C^a + c_0 DG^a) \wedge (v^b + G^b)\wedge (v^c+G^c) \\ &
 + \tfrac16 D c_0 \wedge \epsilon_{abc} (v^a + G^a) \wedge (v^b +G^b) \wedge (v^c + G^c)   + D c_{aI} \wedge (v^a+G^a) \wedge \omega^I \\ &
 + (g^a_{I}+ c_0 t^a_I + \epsilon^{abc} T^J_{bI} c_{cJ} + t^{(ab)} c_{bI})  \tfrac12 \epsilon_{ade} (v^d + G^d)\wedge (v^e+G^e) \wedge \omega^I \\ &
 + (g_0+ c_{aI} t^a_J \eta^{IJ}) \vol^{(0)}_4 \ ,
\end{aligned} \end{equation}
where we defined $DG^a$ in \eqref{eq:con_exp} and the other covariant derivatives are
\begin{equation}\label{eq:covC3} \begin{aligned}
D \hat C_a = & \diff \hat C_a - \epsilon_{abc} t^{db} G^c \wedge \hat C_d \ , \\
D C_I = & \diff C_I +t^a_I \hat C_a - T^J_{aI} G^a \wedge C_J + \tfrac12 \epsilon_{abc} t^{bc} G^a \wedge C_I + \tfrac12 g^a_{I} \epsilon_{abc} G^b \wedge G^c \ , \\
D C^a = & \diff C^a + t^{ba} \hat C_b  + t^{ab} \epsilon_{bcd}  G^c \wedge C^d - \epsilon_{bcd} t^{cd} G^b \wedge C^a \ , \\
D c_0 = & \diff c_0 - \epsilon_{abc} t^{bc} (C^a + c_0  G^a) \ , \\
D c_{aI} = & \diff c_{aI} - T^J_{aI} C_J + \tfrac12 \epsilon_{abc} t^{bc} C_I + \epsilon_{abc} t^b_I C^c - \epsilon_{abc} (g^b_{I}+ t^{db} c_{dI}) G^c   \ .
\end{aligned} \end{equation}

Now we can insert this into \eqref{eq:actionG} and perform the Weyl rescaling \eqref{eq:Weyl}. We find for the kinetic term
\begin{equation} \label{eq:actionGkin} \begin{aligned}
S_{\rm kin} = & \ -\tfrac{1}{4\kappa_{4}^2}  \int_{4} ( \e^{3 \phi+ \rho_3}  (\diff \hat C + \hat C_a \wedge DG^a )\wedge \ast_4 (\diff \hat C + \hat C_b \wedge DG^b ) \\ &
 - \e^{2\phi } g_3^{ab} (D \hat C_a + \epsilon_{acd} C^c \wedge DG^d) \wedge \ast_4 (D \hat C_b + \epsilon_{bef} C^e \wedge DG^f) \\ &
 + \e^{\rho_3}H^{IJ}  (DC_I + c_{aI} DG^a) \wedge \ast_4 (DC_J + c_{bJ} DG^b)\\ &
 + \e^{\phi-\rho_3} g_{3\, ab}(D C^a + c_0 DG^a)\wedge \ast_4 (D C^b + c_0 DG^b) \\ &
- \e^{-2\rho_3} D c_0 \wedge \ast_4 D c_0 - \e^{- \phi} H^{IJ} g_3^{ab} D c_{aI} \wedge \ast_4 D c_{bJ} \\ &
 + \e^{- 2\phi} (g^a_{I}+ c_0 t^a_I + \epsilon^{acd} T^K_{cI} c_{dK} + t^{(ac)} c_{cI}) H^{IJ} g_{3\, ab}\\ &
 \qquad (g^b_{J}+ c_0 t^b_J + \epsilon^{bef} T^L_{eJ} c_{fL} + t^{(be)} c_{eI}) \\ &
 + \e^{-3\phi +\rho_3} |g_0+ c_{aI} t^a_J \eta^{IJ}|^2 )
 \ .
\end{aligned} \end{equation}

To evaluate the topological term correctly in the presence of four-form flux, we assume eleven-dimensional spacetime to be the boundary of a fictional twelve-dimensional space and write the topological term of \eqref{eq:actionG} as \cite{Witten:1996md,Kachru:2004jr}
\begin{equation}\begin{aligned}
 S_{\rm top} = & -\tfrac{1}{12\kappa_{11}^2} \int_{12} G_4 \wedge G_4 \wedge G_4 \\ =&
  -\tfrac{1}{4\kappa_{11}^2} \int_{11} G^{\rm flux}_4 \wedge \diff C^T \wedge C^T  -\tfrac{1}{4\kappa_{11}^2} \int_{11} G_4 \wedge G_4 \wedge C^V  \\ & +\tfrac{1}{4\kappa_{11}^2} \int_{11} G_4 \wedge \diff C^V \wedge C^V -\tfrac{1}{12\kappa_{11}^2} \int_{11} \diff C^V \wedge \diff C^V \wedge C^V \ ,
\end{aligned} \end{equation}
where we used that the flux $G_4^{\rm flux}$ squares to zero, cf.\ \eqref{eq:flux}, and defined $C^T$ to be the part of $C_3$ with two or more external legs, while $C^V$ is the component of $C_3$ with one or less external leg. In other words,
\begin{equation} \begin{aligned}
C^V = & C_I \wedge \omega^I + \tfrac12 \epsilon_{abc} C^a \wedge (v^b +G^b) \wedge (v^c + G^c) +c_{aI} (v^a + G^a) \wedge \omega^I \\ &+ \tfrac16  c_{0} \epsilon_{abc} (v^a + G^a) \wedge (v^b +G^b) \wedge (v^c + G^c) \ ,
\end{aligned}\end{equation}
and
\begin{equation}\begin{aligned}
  -\tfrac{1}{4\kappa_{11}^2} \int_{11} G^{\rm flux}_4 \wedge \diff C^T \wedge C^T = -\tfrac{1}{4\kappa_{4}^2}  \int_{4} (g_0 t^{ab} + g^a_I \eta^{IJ} t^b_J ) \hat C_a \wedge \hat C_b \ .
\end{aligned}\end{equation}

Now we integrate out the three-form $\hat C$, or, more easily, integrate out its field strength $\hat F= \diff \hat C + \hat C_a \wedge DG^a $, since $\hat C$ does not appear by itself in the action. The equation of motion for $\hat F$ is
\begin{equation} \label{eq:eomhatF}
\hat F = - \e^{-3 \phi- \rho_3} (f_0 + (g_0+ c_{aI} t^a_J \eta^{IJ}) c_0 +  (g^a_{I}+ \tfrac12  \epsilon^{abc} T^J_{bI} c_{cJ} + \tfrac12 t^{(ab)}  c_{bI})\eta^{IK} c_{aK} )(\ast_4 1) \ ,
\end{equation}
and inserting this for $\hat F$ in the action leads to an additional term for the potential
\begin{equation}
 S_{\hat F} = \tfrac{1}{4\kappa_{4}^2}  \int_{4} (\ast_4 1) \e^{-3 \phi- \rho_3}|(g_0+ c_{aI} t^a_J \eta^{IJ}) c_0 +  (g^a_{I}+ \tfrac12  \epsilon^{abc} T^J_{bI} c_{cJ} + \tfrac12 t^{(ab)} c_{bI})\eta^{IJ} c_{aJ} |^2 \ .
\end{equation}

Note that the topological term then reduces to
\begin{equation} \label{eq:actionGtop}\begin{aligned}
S_{\rm top} =  -\tfrac{1}{4\kappa_{11}^2} \int_4 ( & c_0 \eta^{IJ} (DC_I + c_{aI} DG^a) \wedge (DC_J + c_{bJ} DG^b) \\ &
+  (D \hat C_a - \epsilon_{ade} C^d \wedge DG^e)  \\ &\qquad \wedge (\epsilon^{abc} c_{cI} \eta^{IJ} Dc_{bJ}+ 2 f_0 G^a + 2 g_0 C^a + 2 g_I^a \eta^{IJ} C_J) \\ &
+ c_{aI} \eta^{IJ} (2DC_J + c_{bJ} DG^b) \wedge D C^a ) + (g_0 t^{ab} + g^a_I \eta^{IJ} t^b_J ) \hat C_a \wedge \hat C_b \\ &
+ \eta^{IJ} (DC_I - t^a_I \hat C_a) \wedge C^b \wedge (\tfrac12 \epsilon_{bcd} t^c_J C^d - T^K_{bJ} C_K + \tfrac12 \epsilon_{bcd} t^{cd} C_J) \\ &
-  ( f_0 G^a + \tfrac12 g_0 C^a)  \wedge \epsilon_{abc} C^b \wedge DG^c \ ,
\end{aligned}\end{equation}
where $S_{\rm top, vec}$ only depends on the vector fields $C_I$ and $C^a$.

Finally, we want to introduce scalar fields $\gamma^a$ so that the kinetic term of the $\hat C_a$ can be replaced. To be consistent, we also have to introduce magnetic vector fields $\tilde C_I$ and $\tilde C_a$ that are dual to $C_I$ and $C^a$, as well as a number of auxiliary two-form fields. Also, we want to perform an electric-magnetic duality between $C^a$ and $\tilde C_a$ to end up in the standard frame of $N=4$ gauged supergravity. Since the scalars $c_{aI}$ and $c_0$ are charged under $C^a$, $\tilde C_I$ and $\tilde C_a$ will be charged under their dual two-forms $\hat C^a_I$ and $\hat C_0$, which must be introduced as well. Note that this very much complicates the situation compared to \cite{Cassani:2012pj,KashaniPoor:2013en}.

When dualizing fields, Bianchi identities and field equations are swapped. This means that from the Bianchi identities of $\hat F_a = D \hat C_a - \epsilon_{abc} C^b \wedge DG^c$ and $F^a = DC^a + c_0 DG^a$ we can deduce the couplings of their dual fields $\gamma^a$ and $\tilde C_a$ in the Lagrangian. On the other hand, the field equations of $\hat F_a$ and $F^a$ tell us what should be the covariant derivatives of $\gamma^a$ and $\tilde C_a$. In particular, we can see that if $C^a$ appears in the covariant derivative of scalar fields, their dual tensors have to appear in the covariant derivative of $\tilde C_a$, and there should be an additional topological coupling of this tensor to $C^a$ in the final Lagrangian. Furthermore, if $\hat C_a$ appears in the covariant derivative of an electric gauge field, the scalar $\gamma^a$ must be gauged under the magnetic dual of the gauge field, and $\hat C_a$ should be topologically coupled to this magnetic vector.

In Appendix \ref{app:dual} we perform the duality transformation from $\hat C_a$ and $C^a$ to $\gamma^a$ and $\tilde C_a$. Both $\hat C_a$ and $C^a$ become auxiliary fields without kinetic terms. Moreover, also a new auxiliary vector field $\tilde C_I$ and the auxiliary tensors $\hat C_0$ and $\hat C^a_I$ appear in the dual Lagrangian.
These dual fields will mostly appear through their covariant derivatives
\begin{equation}\label{eq:covder_dual} \begin{aligned}
 D \hat C_0 = & \diff \hat C_0 + \epsilon_{abc} t^{bc} G^a \wedge C_0 - \epsilon_{abc}t^a_I \eta^{IJ} G^b \wedge \hat C^c_J  - \tfrac12 t^a_I \eta^{IJ} C_J \wedge \hat C_a \\ &
 - f_0 \epsilon_{abc} G^a \wedge G^b \wedge G^c \ , \\
 D\hat C^a_I = & \diff \hat C^a_I - T^J_{bI} G^a \wedge \hat C^b_J - \tfrac12 \epsilon_{bcd} t^{cd} G^a \wedge \hat C^b_I -  t^{ba} C_I \wedge \hat C_b   \ , \\
D \tilde C_a = & \diff \tilde C_a - \epsilon_{abc} t^{bc} \hat C_0 - \epsilon_{abc} t^b_I \eta^{IJ} \hat C^c_J - g_0 \hat C_a - \tfrac12 T^I_{aK} \eta^{KJ} C_I \wedge C_J \\ &
 + \epsilon_{bcd} t^{bc} G^d \wedge\tilde C_a  - \epsilon_{abc} t^{db} G^c \wedge \tilde C_d +\tfrac12 f_0 \epsilon_{abc} G^b \wedge G^c \ , \\
D\tilde C_I = & \diff \tilde C_I + T^J_{aI} \hat C^a_J + \tfrac12 \epsilon_{abc} t^{bc} \hat C^a_I - g^a_I \hat C_a + \tfrac12 t^a_I \epsilon_{abc} C^b \wedge C^c- g^a_I \epsilon_{abc} G^b \wedge C^c \\ &
+ T^J_{aI} C^a \wedge C_J
+ \tfrac12 \epsilon_{abc}t^{bc} C^a \wedge C_I  - T^J_{aI} G^a \wedge \tilde C_J - \tfrac12 \epsilon_{abc} t^{bc} G^a \wedge \tilde C_I
\ , \\
D \gamma^a = & \diff \gamma^a - (f_0 \delta^a_d + t^{ab} \epsilon_{bcd} (\gamma^c + \tfrac12 \epsilon^{cef} c_{eI} \eta^{IJ} c_{fJ})) G^d - t^{ab} \tilde C_b - t^a_I \eta^{IJ} \tilde C_J - g_0 C^a \\ &
- g^a_I \eta^{IJ} C_J
 + c_{bI} \eta^{IJ} ( t^{[a}_J C^{b]} - g^{[a}_J G^{b]} )
+ \tfrac12 \epsilon^{abc}c_{bI} \eta^{IJ} T^K_{cJ} C_K + \tfrac12 t^{[ab]} c_{bI} \eta^{IJ} C_J   \ .
\end{aligned} \end{equation}
The kinetic terms in the dual Lagrangian can be computed to
\begin{equation} \label{eq:actionGkindual} \begin{aligned}
\tilde S_{\rm kin} = & \ -\tfrac{1}{4\kappa_{4}^2}  \int_{4} \Big( \e^{\rho_3}H^{IJ}  (DC_I + c_{aI} DG^a) \wedge \ast_4 (DC_J + c_{bJ} DG^b)\\ &
+\e^{\rho_3-\phi} g^{ab}_{3}( D \tilde C_a - c_{aI} \eta^{IK} DC_K - \epsilon_{acd} \gamma^c DG^d -\tfrac12 c_{aI} \eta^{IK} c_{cK} DG^c) \\ &
\qquad \wedge  \ast (D \tilde C_b - c_{bJ} \eta^{JL} DC_L - \epsilon_{bef} \gamma^e DG^f -\tfrac12 c_{bJ} \eta^{JL} c_{eL} DG^e)  \\ &
- \e^{-2\rho_3} D c_0 \wedge \ast_4 D c_0 - \e^{- \phi} H^{IJ} g_3^{ab} D c_{aI} \wedge \ast_4 D c_{bJ} \\ &
- \e^{-2\phi } g_{3\, ab} (D \gamma^a +\tfrac12 \epsilon^{acd} c_{cI} \eta^{IJ} Dc_{dJ}) \wedge \ast_4 (D \gamma^b+\tfrac12 \epsilon^{bef} c_{eI} \eta^{IJ} Dc_{fJ}) \Big)
 \ ,
\end{aligned} \end{equation}
the potential is given by
\begin{equation} \label{eq:actionGpotdual} \begin{aligned}
\tilde S_{\rm pot} = & \ -\tfrac{1}{4\kappa_{4}^2}  \int_{4} \Big( \e^{- 2\phi} (g^a_{I}+ c_0 t^a_I + \epsilon^{acd} T^K_{cI} c_{dK} + t^{(ac)} c_{cI}) H^{IJ} g_{3\, ab}\\ &
 \qquad (g^b_{J}+ c_0 t^b_J + \epsilon^{bef} T^L_{eJ} c_{fL} + t^{(be)} c_{eI}) \\ &
+ \e^{-3 \phi- \rho_3}|f_0+ (g_0+ c_{aI} t^a_J \eta^{IJ}) c_0 +  (g^a_{I}+ \tfrac12  \epsilon^{abc} T^J_{bI} c_{cJ} + \tfrac12 t^{(ab)} c_{bI})\eta^{IJ} c_{aJ} |^2 \\ &
 + \e^{-3\phi +\rho_3} |g_0+ c_{aI} t^a_J \eta^{IJ}|^2 \Big)
 \ ,
\end{aligned} \end{equation}
and the topological terms are
\begin{equation} \label{eq:actionGtopdual} \begin{aligned}
\tilde S_{\rm top} = & \ -\tfrac{1}{4\kappa_{4}^2}  \int_{4} \Big(c_0 (\eta^{IJ}DC_I \wedge DC_J - 2 D \tilde C_a \wedge D G^a) \\ & + g_0 (2DC^a - t^{ba} \hat C_b) \wedge \hat C_a+ 2 \epsilon_{abc} t^a_I \eta^{IJ} \tilde C_J \wedge C^b \wedge DG^c
\\ & +  2 DC^a \wedge \epsilon_{abc} (t^{bc}\hat C_0 + t^b_I \eta^{IJ} \hat C^c_J) - t^a_I\eta^{IJ} (2D \tilde C_J - g^b_J \hat C_b) \wedge \hat C_a
\\ & + \epsilon_{abc} t^{bc} \eta^{IJ} C^a \wedge C_I \wedge DC_J + t^a_I \eta^{IJ} \epsilon_{abc} C^a \wedge C^b \wedge D C_J \\ &
+ \tfrac16 f_0 \epsilon_{def} t^{ef} \epsilon_{abc} G^a \wedge G^b \wedge G^c \wedge C^d
\Big) \ .
\end{aligned} \end{equation}
Variation with respect to the auxiliary tensor fields leads to the duality relations between electric and magnetic vectors
\begin{equation}\label{eq:dual_rel_vectors}\begin{aligned}
D C_a+ c_0 DG^a = & \e^{\rho_3-\phi} g_{3}^{ab} \ast  (D \tilde C_b- c_{bI} \eta^{IK} DC_K - \epsilon_{bcd} \gamma^c DG^d -\tfrac12 c_{bI} \eta^{IK} c_{cK} DG^c)\ , \\
D\tilde C_I - c_{aI} DC^a   = & \e^{\rho_3} H^J_I \ast (DC_J +c_{aI} DG^a) + c_0 (DC_I+c_{aI} DG^a) \ ,
\end{aligned} \end{equation}
while variation with respect to the magnetic vectors gives the duality relations between tensors and scalars
\begin{equation}\label{eq:dual_rel_tensors}\begin{aligned}
D \hat C_a -\epsilon_{abc} C^b \wedge DG^c = & \e^{- 2\phi} g_{3\,ab} \ast (D\gamma^b +\tfrac12 \epsilon^{acd} c_{cI} \eta^{IJ} Dc_{dJ})\ , \\
D \hat C_0 + \tfrac12 \eta^{IJ} C_I \wedge F_J = & \e^{-2\rho_3} \ast Dc_0 \ , \\
D\hat C^a_I - C^a \wedge DC_I - c_{bI} C^b \wedge DG^a - \epsilon^{abc} c_{bI} D\hat C_c= & \e^{-\phi} H_I^J g^{ab}_3 \ast Dc_{bJ} \ .
\end{aligned} \end{equation}
Note that each of the relations in \eqref{eq:dual_rel_tensors} and \eqref{eq:dual_rel_vectors} gets multiplied by certain charge components. Thus, if certain charges are vanishing, the corresponding duality equation is eliminated. At the same time, the corresponding couplings in the covariant derivative vanish and the corresponding auxiliary field is removed from the Lagrangian altogether.
In the generic case of non-vanishing couplings, we can use the duality relations \eqref{eq:dual_rel_tensors} and \eqref{eq:dual_rel_vectors} in order to eliminate the fields we have introduced above and come back to the Lagrangian of \eqref{eq:actionGkin} and \eqref{eq:actionGtop} that we obtained from the reduction.

The action obtained in \eqref{eq:actionGkindual}, \eqref{eq:actionGpotdual} and \eqref{eq:actionGtopdual} together with \eqref{eq:actiongrav} fits perfectly into the framework of $N=4$ gauged supergravity. We make the identification with the standard notation in the next section.

\section{Matching with $N=4$ supergravity}
\label{sec:matching}

In this section we want to match the results of the dimensional reduction on a seven-dimensional SU(2)-structure manifold with the standard formulation of $N=4$ gauged supergravity, which is reviewed in Appendix~\ref{sec:N=4gaugedSUGRA}.
We organize the vector fields as
\begin{equation}
V^{M+} = (G^a, \tilde C_a, \eta^{IJ} C_J ) \ , \qquad V^{M-} = (C^a, \tilde G_a , - \eta^{IJ} \tilde C_J) \ ,
\end{equation}
so that the metric $\eta_{MN}$ is in the standard form
\begin{equation}
 \eta_{MN} = \left( \begin{aligned}  0 && -\delta_a^b && 0 \\ -\delta^a_b && 0  && 0 \\ 0 && 0 && \eta_{IJ} \end{aligned} \right) \ .
\end{equation}
Note that this involves an electric-magnetic duality transformation between $\tilde C_a$ and $C^a$, which can be performed in the standard way following \cite{Schon:2006kz}.

Next, the scalars $c_0$ and $\rho_3$ combine into the $N=4$ axiodilaton $\tau$ as $\tau = (-c_0 + \iu \e^{\rho_3} )$ so that the covariant derivative reads
\begin{equation}
 D \tau = \diff \tau - \epsilon_{abc} t^{bc} (C^a + \tau G^a) \ .
\end{equation}
Therefore we find for the embedding tensor components $\xi_{\alpha M}$ that the only non-vanishing component is
\begin{equation} \label{eq:embeddingtensor1}
 \xi_{+a} = - \epsilon_{abc} t^{bc} \ .
\end{equation}

The coset matrix $M_{MN}$ of the Grassmanian
\begin{equation}
 \frac{SO(6,n)}{SO(6) \times SO(n)}
\end{equation}
is given in terms of the scalars $(\phi, g_{3\,ab}, \zeta^{\hat a}_I, \gamma^a, c_{aI})$ by
\begin{equation}\begin{aligned}
M_{ab} = & \e^{\phi} g_{3\, ab} - \tfrac14 \e^{-\phi} g_3^{cd} (\epsilon_{ace} \gamma^e + c_{aI} \eta^{IJ} c_{cJ}) (\epsilon_{bdf} \gamma^f + c_{bK} \eta^{KL} c_{dL}) + H^{IJ} c_{aI} c_{bJ} \ , \\
M^a_b = & \tfrac12 \e^{-\phi}  g_3^{ac} ( \epsilon_{cbd} \gamma^d + c_{cI} \eta^{IJ} c_{bJ})\ , \\
M^{ab} = & \e^{-\phi} g_3^{ab} \ , \\
M_{aI} = & H^J_I c_{aJ} + \tfrac12 \e^{-\phi} (\epsilon_{abc} \gamma^b + c_{aI} \eta^{IJ} c_{bJ} ) g_3^{cd} c_{dI} \ , \\
M^a_I = & \e^{-\phi} g_3^{ab} c_{bI} \ , \\
M_{IJ} = & H_{IJ} + \e^{-\phi} g_3^{ab} c_{aI} c_{bJ} \ .
\end{aligned}\end{equation}
The corresponding vielbein $\mathcal{V}$ is given by
\begin{equation} \label{eq:vielbeinSUGRA}\begin{aligned}
\mathcal{V}^b{}_a = & \e^{\phi/2} k^b_a + \e^{-\phi/2} (k^{-1})_b^c (\epsilon_{cad} \gamma^d + c_{cI} \eta^{IJ} c_{aJ} ) \ , \\
\mathcal{V}^b{}^a = & \e^{-\phi/2} (k^{-1})_b^a \ , \\
\mathcal{V}^b{}_I = & \e^{-\phi/2} (k^{-1})_b^c c_{cI} \ , \\
\mathcal{V}^{\hat b}{}_a = &\zeta^{\hat b I} c_{a I} \ , \\
\mathcal{V}^{\hat b}{}^a = & 0 \ , \\
\mathcal{V}^{\hat b}{}_I = & \zeta^{\hat b}_I \ ,
\end{aligned}\end{equation}
so that
\begin{equation}
 M_{MN} = 2 \mathcal{V}_M^{a} \mathcal{V}_N^{a} + 2 \mathcal{V}_M^{\hat a} \mathcal{V}_N^{\hat a} - \eta_{MN} \ .
\end{equation}

From the covariant derivatives we can read off the remaining embedding tensor components
\begin{equation}\label{eq:embeddingtensor2}\begin{aligned}
f_{+abc}  = & f_0 \epsilon_{abc}    \ , \\
f_{+ab}{}^c  = &  - \epsilon_{abd} t^{cd}    \ , \\
f_{+aIJ} = & \eta_{IK} T^K_{aJ} \ ,  \\
f_{+abI} = & \epsilon_{abc} g^c_I \ , \\
f_{-abc}  = & g_0 \epsilon_{abc}    \ , \\
f_{-abI} = & \epsilon_{abc} t^c_I \ .
\end{aligned}\end{equation}

Moreover, we identify
\begin{equation}\begin{aligned}
B^{++} = & 2 \hat C_0\ , \\
B^{ab} = & \epsilon^{abc} \hat C_c + C^a \wedge G^b \ , \\
B^{aI} = & \eta^{IJ} (2 \hat C^a_J + C^a \wedge C_J - G^a \wedge \tilde C_J )\ .
\end{aligned}\end{equation}
Comparing the charges \eqref{eq:embeddingtensor1} and \eqref{eq:embeddingtensor2} with \cite{Schon:2006kz}, we see that the remaining auxiliary two-forms as well as the magnetic vector $\tilde G_a$ do not explicitly appear in the Lagrangian.
Moreover one can check that the constraints \eqref{eq:quadratic_constraints}, \eqref{eq:quadrem} and \eqref{eq:quadflux} solve the quadratic constraints \eqref{eq:quad_constr}.

\section{Consistent truncation}
\label{sec:consistent}

In this section we want to show that the $SU(2)$ reduction to four dimensions is indeed a consistent truncation of the eleven-dimensional supergravity action. In other words, we want to show that the four-dimensional equations of motion imply the ten-dimensional ones.
In \cite{KashaniPoor:2013en} a heuristic argument was already given why $SU(2)$ structure reductions for modes $\omega^I$ and $v^a$ that obey the constraints \eqref{eq:omegavol4} and \eqref{eq:truncation_ansatz} are consistent truncations. In this section we will prove this claim by an explicit check of the eleven-dimensional equations of motion.

The four-dimensional relevant equations are the Einstein equation
\begin{equation}\label{eq:eom4d0} \begin{aligned}
\hat R_{\mu\nu}  = & - 2 {\rm tr}(((D_\mu \mathcal{V}) \mathcal{V}^{-1} + (\mathcal{V}^{-1})^T (D\mathcal{V})^T)^2((D_\nu \mathcal{V}) \mathcal{V}^{-1} + (\mathcal{V}^{-1})^T (D\mathcal{V})^T)) \\ &
-\tfrac12 (\Im \tau)^{-2} D_{(\mu} \tau D_{\nu)} \bar \tau - \Im(\tau) M_{MN} DV^{M+}_{\mu \rho} DV^{N+}_{\nu}{}^\rho +  \tfrac12 g_{\mu\nu} ({\cal L}_{\rm kin} + {\cal L}_{\rm pot})   \ ,
\end{aligned}\end{equation}
the equations of motion for the vector fields
\begin{equation}\label{eq:eom4d1} \begin{aligned}
D( \Re (\tau) DV^{M+} + \Im (\tau) M^{MP} \eta_{PN} DV^{N+} ) = & - \eta^{MN} (\xi_{\beta M} M_{+\gamma} \ast_4 DM^{\beta \gamma} \\ & + \tfrac12 \theta_{MP}{}^{N} M_{NQ} \ast_4 DM^{QP} ) \ ,
\end{aligned}\end{equation}
and for the scalars
\begin{equation}\label{eq:eom4d2} \begin{aligned}
&  D  ( (\mathcal{V}^{-1})^n_M \ast_4 ( (D\mathcal{V}) \mathcal{V}^{-1} + (\mathcal{V}^{-1})^T (D\mathcal{V})^T)_{mn}) - \tfrac14 \Im (\tau )\mathcal{V}^m_N  DV^{M+}\wedge \ast_4 DV^{N+}  \\ = & \ \tfrac{1}{16} ( M^{\alpha \beta} f_{\alpha MPR} f_{\beta NQS} (\mathcal{V}^{-1})^N_m (M^{PQ} M^{RS} - \eta^{PQ} \eta^{RS}) + 3 M^{\alpha \beta} \mathcal{V}^m_N \xi_\alpha^M \xi_\beta^N) \\ & - \tfrac23 \epsilon^{\alpha \beta} f_{\alpha \tilde M NP} f_{\beta QRS} \mathcal{V}_M^m  M^{MNPQRS} \ , \\
 & D^\mu ( (\Im \tau)^{-2} D_\mu \tau) - \tfrac14 ( \eta_{MN}  + \iu M_{MN} \ast_4) DV^{M+} \wedge DV^{N+} \\ & = \tfrac18 (\partial_{\bar \tau} M^{\alpha \beta}) (f_{\alpha MNP} f_{\beta QRS}( \tfrac13 M^{MQ} M^{NR}M^{PS}  +(\tfrac23 \eta^{MQ} - M^{MQ}) \eta^{NR} \eta^{PS} ) \\ &\hskip2.5cm  + 3 \xi^M_\alpha \xi^N_\beta M_{MN}) \ ,
\end{aligned}\end{equation}
as well as the identities for the auxiliary field strengths\footnote{The covariant derivative of $DV^+$ gives via the Bianchi identity the three-form field strengths of the auxiliary tensor fields, cf.~\cite{Schon:2006kz}.}
\begin{equation}\label{eq:idaux4d} \begin{aligned}
DV^{M-} = & - \Re (\tau) DV^{M+} - \Im (\tau) M^{MP} \eta_{PN} \ast_4 DV^{N+} \ , \\
\eta_{MN} \ast_4 D DV^{N+} = & -\xi_{\alpha M} M_{-\beta} DM^{\alpha \beta} - \tfrac12 \Theta_{MP}{}^{N} M_{NQ} D M^{QP} \ .
\end{aligned}\end{equation}
originating from the Lagrangian of \cite{Schon:2006kz} discussed in Appendix~\ref{sec:N=4gaugedSUGRA}.
The eleven-dimensional equations of motion originate from the action \eqref{eq:actionM} and are given by
\begin{equation}\label{eq:eom11d} \begin{aligned}
{\rm Ric}_{AB} - \tfrac12 g_{AB} r_{11} = & \tfrac{1}{12} ( G_{ACDE} G_B{}^{CDE} - \tfrac14 g_{AB} G_{CDEF} G^{CDEF}) \ , \\
\diff \ast_{11} G_4 = & - \tfrac12 G_4 \wedge G_4 \ .
\end{aligned} \end{equation}

The major work consists of showing that the eleven-dimensional Einstein equations are satisfied if the four-dimensional equations of motion (and Bianchi identities) hold.
The technical details for determining the Ricci curvature and the energy-momentum tensor are delegated to the first two appendices.
In Appendix~\ref{app:RicEinstein} we give the Ricci curvature in the Einstein frame. In Appendix~\ref{app:emtensor}, we also compute the energy-momentum tensor generated by $G_4$. When we insert these results into the eleven-dimensional Einstein equation, we see that the equations reduce to the four-dimensional equations of motion in the following way:
\begin{itemize}
\item The trace of the Einstein equations is satisfied by the equations of motion for $\phi$, $\rho_3$ and by the trace of the four-dimensional Einstein equation.
\item The Einstein equations with indices $(\mu\nu)$ give the four-dimensional Einstein equations.
\item For the indices $(\mu a)$ we recover the equations of motion for the Kaluza-Klein vector $G^a_\mu$.
\item The trace of the $(ab)$ component of the eleven-dimensional Einstein equations gives the equation of motion for $\rho_3$, while the traceless part is the equation of motion for $k^a_b$.
\item The trace of the $(\alpha \beta)$ component of the eleven-dimensional Einstein equations is the equation of motion for $\rho_4$, i.e.\ for $\phi - 2 \rho_3/3$. The traceless part is the equation of motion for $\zeta^{\hat a}_I$.
\end{itemize}

For the higher form field components we then use the four-dimensional equation of motion for $\hat F$ \eqref{eq:eomhatF} to eliminate $\hat F$, the scalar-tensor duality relation \eqref{eq:dual_rel_tensors} to replace the tensor fields by their dual scalars and the electro-magnetic duality relation \eqref{eq:dual_rel_vectors} to replace all magnetic vector fields by their electric counterparts. In this way we can rewrite the eleven-dimensional Einstein equations in terms of four-dimensional scalars and electric vector fields (up to appearances of the magnetic vectors and tensors in the gaugings). As expected, this completely reproduces the four-dimensional equations of motions for these scalars and vector fields.

\section{Simple supersymmetric backgrounds}
\label{sec:vacua}

Let us now briefly discuss some classes of supersymmetric AdS vacua. We will only discuss the simple examples of $N=4$ AdS vacua discussed in \cite{Louis:2014gxa} and of $N=3$ AdS vacua from Tri-Sasakian manifolds whose consistent truncation has been worked out in \cite{Cassani:2011fu}. The discussion of cases with $N \le 2$ goes beyond the scope of this paper.

\subsection{$N=4$ AdS vacua}
In \cite{Louis:2014gxa} four-dimensional $N=4$ AdS vacua had been classified. A necessary requirement for such backgrounds is that there is one electrically and one magnetically gauged $SU(2)$ in the theory whose gauge bosons are graviphotons. Let us apply the findings of \cite{Louis:2014gxa} to the gaugings of $SU(2)$ structures given in \eqref{eq:embeddingtensor1} and \eqref{eq:embeddingtensor2}. The embedding tensor component $\xi$ in \eqref{eq:embeddingtensor1} must be zero in these vacua, which means that $t^{ab}$ is symmetric. Moreover, the electric and magnetic gaugings obey the relationship
\begin{equation}
f_+ = \Re \tau f_- + \Im \tau \ast_6 f_- \ ,
\end{equation}
where $\ast_6$ is the Hodge star in the six-dimensional space of graviphotons and $\tau$ is the axiodilaton. For the possible gaugings given in \eqref{eq:embeddingtensor2}, this means that
\begin{equation} \label{eq:AdSN4a}
\mathcal{V}^{a d} \mathcal{V}^{\hat b I} \mathcal{V}^{\hat c J } f_{+dIJ} \ne 0 \ ,
\end{equation}
where the supergravity vielbein has been given in \eqref{eq:vielbeinSUGRA}.
But $N=4$ supersymmetry also requires that
\begin{equation}\label{eq:AdSN4b}
\mathcal{\tilde V}^{\tilde a d} \mathcal{V}^{\hat b I} \mathcal{V}^{\hat c J } f_{+dIJ} = 0 \ ,
\end{equation}
where the dual vielbein $\mathcal{\tilde V}^{\tilde a}$ is given by
\begin{equation} \begin{aligned}
\mathcal{\tilde V}^{\tilde b}{}_a = & - \e^{\phi/2} k^{\tilde b}_a - \e^{-\phi/2} (k^{-1})_{\tilde b}^c (\epsilon_{cad} \gamma^d + c_{cI} \eta^{IJ} c_{aJ} ) \ , \\
\mathcal{\tilde V}^{\tilde b}{}^a = & \e^{-\phi/2} (k^{-1})_{\tilde b}^a \ , \\
\mathcal{\tilde V}^{\tilde b}{}_I = & \e^{-\phi/2} (k^{-1})_{\tilde b}^c c_{cI} \ .
\end{aligned} \end{equation}
Comparing the two formulas \eqref{eq:AdSN4a} and \eqref{eq:AdSN4b}, using \eqref{eq:vielbeinSUGRA} and the above formula, shows the contradiction. Thus, in the class of supergravities obtained from $SU(2)$ structure truncations no $N=4$ AdS vacua can be found.

\subsection{Tri-Sasakian manifolds}
The consistent truncations worked out in \cite{Cassani:2011fu} for Tri-Sasakian manifolds admit $N=3$ AdS vacua. These truncations are minimal in that $I$ runs $I=1,2,3$. There the ansatz is
\begin{equation}
 g_0 = 0 \ , \qquad g^a_I = 0\ , \qquad t^{ab} = -2 \delta^{ab} \ , \qquad t^a_I = 2 \delta^a_I \ , \qquad T^I_{aJ} = - 2 \epsilon_{aIJ} \ .
\end{equation}
A classification of $N=3$ vacua is beyond the scope of this paper, but note that one could easily for instance add a non-trivial four-form flux by switching on $g^a_I \sim \delta^a_I$. These four-form fluxes effectively just change the value of the axiodilaton $\tau$ but do not modify the solution in any other way.

An interesting special case is the reduction on $S^7$. We have already shown that the discussed $SU(2)$ structure reductions do not allow for $N=4$ AdS vacua. And indeed, the consistent truncation presented here does not distinguish between $S^7$ and any other Tri-Sasakian manifold. Thus while the full vacuum of $AdS_4 \times S^7$ preserves $N=8$ supersymmetry, the truncation to this $N=4$ gauged supergravity (i.e.\  the truncation to $SU(2)$ singlet and triplet modes) preserves only $N=3$.

\section*{Acknowledgments}
We would like to thank A.\ Kashani-Poor and R.\ Minasian for their motivational support as well as for useful discussions and for comments on the final draft.

\appendix

\section{Curvature computations} \label{app:curvature}

\subsection{Useful formulas}

From \eqref{eq:k_det} we find the identities
\begin{equation}
(k^{-1})^b_a = \tfrac12 \epsilon_{acd} \epsilon^{bef} k^c_e k^d_f\ .
\end{equation}
It also implies
\begin{equation}
 (k^{-1})^b_a \diff k^{a}_b  = 0 \ , \qquad  k^a_d \epsilon_{abc} =  \epsilon_{def} (k^{-1})^e_b (k^{-1})^f_c \ .
\end{equation}

From the algebra \eqref{eq:truncation_ansatz} we find
\begin{equation}
\diff (\epsilon_{abc} v^b \wedge v^c) = 2 \epsilon_{abc} t^{bc} {\rm vol}^{(0)}_3 + 2 \epsilon_{abc} t^b_I v^c\wedge \omega^I \ ,
\end{equation}
and
\begin{equation}
\diff {\rm vol}^{(0)}_4 = - \epsilon_{abc} t^{bc} v^a \wedge {\rm vol}^{(0)}_4 \ .
\end{equation}
Another useful formula will be
\begin{equation}
\diff (v^a + G^a) = DG^a- t^{ab} \epsilon_{bcd} G^c \wedge (v^d + G^d) + \tfrac12 t^{ab} \epsilon_{bcd} (v^c + G^c) \wedge (v^d + G^d) + t^a_I \omega^I \ ,
\end{equation}
where we defined the covariant derivative
\begin{equation} \label{eq:covG}
D G^a = \diff G^a + \tfrac12 t^{ab} \epsilon_{bcd} G^c \wedge G^d \ ,
\end{equation}
and
\begin{equation}
\diff \omega^I = -T^I_{aJ} G^a \wedge \omega^J + \tfrac12 \epsilon_{abc} t^{bc} G^a \wedge \omega^I + T^I_{aJ} (v^a + G^a) \wedge \omega^J - \tfrac12 \epsilon_{abc} t^{bc} (v^a + G^a) \wedge \omega^I \ .
\end{equation}
From \eqref{eq:truncation_ansatz} and \eqref{eq:omegavol4} we also deduce that
\begin{equation}
\tilde T^I_{a J} \zeta_I^{(\hat a} \zeta^{\hat b)\,  J} = -\tfrac12 \epsilon_{abc} t^{bc} \delta^{\hat a \hat b} \ .
\end{equation}
Note also that we find from \eqref{eq:decomp_T} that
\begin{equation}
 \epsilon^{\hat a \hat b \hat c} \tilde  T^I_{aJ} \zeta^{\hat b}_I \zeta^{\hat c \, J} = \epsilon^{\hat a \hat b \hat c} T^I_{aJ} \zeta^{\hat b}_I \zeta^{\hat c \, J}  \ ,
\end{equation}

Eq.~\eqref{eq:zeta_ab} implies
\begin{equation}\label{eq:delzeta}
(\partial_\mu \zeta^{\hat a}_I) \eta^{IJ} \zeta^{\hat b}_J = 0 \ ,
\end{equation}
which means that over four-dimensional spacetime, the $J^{\hat a}$ do not rotate into each other and therefore really move in $\frac{SO(3,n)}{SO(3) \times SO(n)}$.
Note that from \eqref{eq:Jexpl} and \eqref{eq:param_forms} we also find
\begin{equation}
( I^{\hat a})^\alpha_\beta = \e^{\rho_4/2} \zeta^{\hat a}_I \omega^I_{\alpha \beta} \ .
\end{equation}

The decomposition of $\omega^I$ into representations of $SU(2)$ reads
\begin{equation} \label{eq:omega_decomp}
\omega^I = \zeta^{\hat a\, I} \zeta^{\hat a}_J \omega^J + P^I_J \omega^J \ ,
\end{equation}
where $P^I_J = \delta^I_J - \zeta^{\hat a\, I} \zeta^{\hat a}_J$.
The latter term in \eqref{eq:omega_decomp} is invariant under the $I^{\hat a}$. We hence find
\begin{equation}\label{eq:SU2rot_omega}
 \tfrac12 (I^{\hat a})^\alpha_\beta (\omega^I)_{\beta \gamma} e^\alpha \wedge e^\gamma =  \e^{-\rho_4/2} \epsilon^{\hat a\hat b\hat c} \zeta^{\hat b\, I} J^{\hat c} \ ,
\end{equation}
From \eqref{eq:delzeta} and from the definition of $P^I_J$ it follows that for the derivative of an $SU(2)$ singlet it holds that
\begin{equation}
 \diff M^I = P^I_J \diff M^J - \zeta^{\hat a\, I} \partial_\mu \zeta^{\hat a}_J M^J e^\mu \ ,
\end{equation}
where we used that $M^I = P^I_J M^J$.

From the definition of the Hodge dual
\begin{equation}
(\ast_4 \omega^I)_{\alpha\beta} = \tfrac12 \epsilon_{\alpha \beta \gamma \delta} \omega^I_{\gamma \delta} = H^I_J \omega^J_{\alpha \beta}\ ,
\end{equation}
and the decomposition \eqref{eq:omega_decomp}
we find
\begin{equation}\label{eq:omega_squared}
 \tfrac14 (\omega^I)_{\alpha \beta }(\omega^J)_{\alpha \beta } = \e^{-\rho_4}H^{IJ} \ ,
\end{equation}
which implies
\begin{equation}\label{eq:Iomega}
 (\omega^I)_{\alpha \beta }  (I^{\hat a})^\alpha_\beta  = 4 \e^{-\rho_4/2} \zeta^{\hat a\, I} \ .
\end{equation}
From \eqref{eq:omega_squared} we can deduce
\begin{equation}
 \omega^I_{\gamma (\alpha } \omega^J_{\beta) \gamma} = - \e^{-\rho_4} H^{IJ} \delta_{\alpha \beta} + 2 \e^{-\rho_4/2} \zeta^{\hat a \, (I} P^{J)}_K \omega^K_{\alpha \gamma}
 (I^{\hat a})^\gamma_\beta\ .
\end{equation}

\subsection{Connection}
\label{sec:connection}

In this section we want to compute the connection from the three Maurer-Cartan equations \eqref{eq:MCexplicit}, using \eqref{eq:param_forms} and \eqref{eq:truncation_ansatz}.
From the first equation in \eqref{eq:MCexplicit} we see that both $\lambda^\mu_\alpha (\hat e_\beta)$ and $\gamma^\mu_a(\hat K_b)$ are symmetric in their lower indices.

Now let us first solve the second equation. From the explicit form of $K^a$ in \eqref{eq:param_forms} we find with help of \eqref{eq:truncation_ansatz}
\begin{equation}\label{eq:dK} \begin{aligned}
 \diff K^a= & \tfrac13 D \rho_3 \wedge K^a + (k^{-1})^c_b (D k^a_c) \wedge K^b + \e^{\rho_3/3} k^a_b DG^b \\ & + \tfrac12 \e^{-\rho_3/3}k^a_b t^{bc} k_c^d \epsilon_{def} K^e \wedge K^f + \e^{\rho_3/3} k^a_b t^b_I \omega^I\ .
\end{aligned}\end{equation}
where the explicit form of $DG^a$ can be found in \eqref{eq:covG} and we defined
\begin{equation} \label{eq:cov_metric1} \begin{aligned}
 D\rho_3 = & \diff \rho_3 - \epsilon_{abc} t^{bc} G^a\ , \\
 D k^a_b = & \diff k^a_b - (k^a_{c} \epsilon_{bde} - \tfrac13 k^a_{b} \epsilon_{cde})  t^{cd} G^e  \ .
\end{aligned}\end{equation}
Comparison with \eqref{eq:MCexplicit} gives for the connection components
\begin{equation}\label{eq:con1} \begin{aligned}
\tau^a_\alpha = & \tfrac12 \e^{\rho_3/3} k^a_b t^b_I \omega^I_{\alpha\beta} e^\beta + \tau^a_0 e^\alpha + \tfrac12 (\tau^{a}_{\hat b})_{\alpha \gamma} (I^{\hat b})^\gamma_\beta e^\beta \ ,\\
\gamma^a_\mu = & \e^{\rho_3/3} k^a_b D_{[\mu} G^b_{\nu]} e^\nu +\gamma^a_{\mu\nu} e^\nu + \tfrac12 [(k^{-1})_b^c  D_\mu k^{a}_c +(k^{-1})_a^c  D_\mu k^{b}_c ] K^b +\tfrac13 (D_\mu \rho_3) K^a \ , \\
\phi^a = & - \e^{-\rho_3/3} (k^a_c k^b_d t^{cd} - \tfrac12 \delta^{ab} k^e_c k^e_d t^{cd}) K^b - \tfrac12 \epsilon^{abc} (k^{-1})_b^d D k_d^c \ ,
\end{aligned}\end{equation}
where $\gamma^a_{\mu \nu}$ is symmetric in its lower indices.
Note that we used \eqref{S2T4representation} for the decomposition of $\tau^a_\alpha$.

If we use the explicit expressions for the $J^{\hat a}$ given in \eqref{eq:param_forms} and \eqref{eq:Jexpl}, the third Maurer-Cartan equation in \eqref{eq:MCexplicit} implies with \eqref{eq:truncation_ansatz}
\begin{equation}\label{eq:MC_Ja} \begin{aligned}
& \tfrac12 \diff \rho_4 \wedge J^{\hat a} + \e^{\rho_4/2} \diff \zeta^{\hat a}_I \wedge \omega^I + \e^{-\rho_3/3} \e^{\rho_4/2} \zeta^{\hat a}_I \tilde T^I_{bJ} (k^{-1})^b_a K^a \wedge \omega^J - \e^{\rho_4/2} \zeta^{\hat a}_I \tilde T^I_{bJ} G^b \wedge \omega^J \\ = & - 2 \epsilon^{\hat a\hat b\hat c} \Psi^{\hat b} \wedge J^{\hat c} + K^a \wedge (I^{\hat a})^\alpha_\beta \tau^\alpha_a \wedge e^\beta + e^\mu \wedge (I^{\hat a})^\alpha_\beta \lambda^\alpha_\mu \wedge e^\beta \ .
\end{aligned}\end{equation}
This can be solved for the connection components as
\begin{equation}\label{eq:con2} \begin{aligned}
\tau^a_\alpha = & \tfrac12 \e^{\rho_3/3} k^a_b t^b_I \omega^I_{\alpha\beta} e^\beta - \tfrac14 \e^{-\rho_3/3} (k^{-1})^b_a \epsilon_{bcd} t^{cd} e^\alpha + \tfrac12 \e^{-\rho_3/3} \e^{\rho_4/2} (k^{-1})^b_a \zeta^{\hat a}_I T^I_{bJ}  P^J_K \omega^K_{\alpha \gamma} ( I^{\hat a})^\gamma_\beta e^\beta \ ,\\
\Psi^{\hat a} = & \tfrac14 \e^{-\rho_3/3} \epsilon^{\hat a \hat b \hat c} \zeta^{\hat b}_I \zeta^{\hat c \, J}  T^I_{bJ} (k^{-1})^b_a K^a - \tfrac14 \e^{\rho_3/3} \e^{-\rho_4/2} \zeta^{\hat a\, I} t^b_I k^a_b K^a - \tfrac14 \epsilon^{\hat a\hat b \hat c} \zeta^{\hat b}_I \zeta^{\hat c\, J} T^I_{bJ} G^b \ , \\
\lambda^\alpha_\mu = & \tfrac14 (D_\mu \rho_4) e^\alpha - \tfrac12 \e^{\rho_4/2} (D_\mu \zeta^{\hat a}_I) \omega^I_{\alpha \gamma} ( I^{\hat a})^\gamma_\beta e^\beta   \ ,
\end{aligned}\end{equation}
where we defined the covariant derivatives
\begin{equation}\label{eq:cov_metric2}\begin{aligned}
  D \rho_4 = & \diff \rho_4 +\epsilon_{abc} t^{bc} G^a   \ , \\
 D \zeta^{\hat a}_I = & \diff \zeta^{\hat a}_I - \zeta^{\hat a}_I T^I_{bJ} (\delta^J_K - \zeta^{\hat b}_K \zeta^{\hat b \, J}) G^b \ .
\end{aligned}\end{equation}
and the projector $P^I_J = (\delta^I_J -  \zeta^{\hat b\, I}\zeta^{\hat b}_J)$.
Finally, we solve for the first Maurer-Cartan equation in \eqref{eq:MCexplicit}, by using the explicit form of $\lambda^\alpha_\mu$ and $\gamma^a_\mu$, given in \eqref{eq:con1} and \eqref{eq:con2}. The result is
\begin{equation}\label{eq:con3} \begin{aligned}
\gamma^a_\mu = & \e^{\rho_3/3} k^a_b D_{[\mu} G^b_{\nu]} e^\nu + \tfrac12 ((k^{-1})_b^c  D_\mu k^{a}_c +(k^{-1})_a^c D_\mu k^{b}_c ) K^b +\tfrac13 (D_\mu \rho_3) K^a \ , \\
 \omega^\mu_\nu = & \hat \omega^\mu_\nu - \e^{\rho_3/3} D_{[\mu} G^b_{\nu]} k^a_b K^a\ ,
\end{aligned}\end{equation}
where $\hat \omega$ is the four-dimensional connection.

In total, we find for the components of the connection
\begin{equation}\label{eq:con_app} \begin{aligned}
\tau^a_\alpha = & \tfrac12 \e^{\rho_3/3} k^a_b t^b_I \omega^I_{\alpha\beta} e^\beta  - \tfrac14 \e^{-\rho_3/3} (k^{-1})^b_a \epsilon_{bcd} t^{cd} e^\alpha  + \tfrac12 \e^{-\rho_3/3} \e^{\rho_4/2} (k^{-1})^b_a \zeta^{\hat a}_I T^I_{bJ}  P^J_K \omega^K_{\alpha \gamma} ( I^{\hat a})^\gamma_\beta e^\beta \ ,\\
\gamma^a_\mu = & \e^{\rho_3/3} k^a_b D_{[\mu} G^b_{\nu]} e^\nu + \tfrac12 ((k^{-1})_b^c  D_\mu k^{a}_c +(k^{-1})_a^c D_\mu k^{b}_c ) K^b +\tfrac13 (D_\mu \rho_3) K^a \ , \\
\phi^a = & - \e^{-\rho_3/3} (k^a_c k^b_d t^{cd} - \tfrac12 \delta^{ab} k^e_c k^e_d t^{cd}) K^b - \tfrac12 \epsilon^{abc} (k^{-1})_b^d D k_d^c \ ,\\
\Psi^{\hat a} = & \tfrac14 \e^{-\rho_3/3} \epsilon^{\hat a \hat b \hat c} \zeta^{\hat b}_I \zeta^{\hat c \, J}  T^I_{bJ} (k^{-1})^b_a K^a - \tfrac14 \e^{\rho_3/3} \e^{-\rho_4/2} \zeta^{\hat a\, I} t^b_I k^a_b K^a - \tfrac14 \epsilon^{\hat a\hat b \hat c} \zeta^{\hat b}_I \zeta^{\hat c\, J} T^I_{bJ} G^b \ , \\
\lambda^\alpha_\mu = & \tfrac14 D_\mu \rho_4 e^\alpha - \tfrac12 \e^{\rho_4/2} D_\mu \zeta^{\hat a}_I \omega^I_{\alpha \gamma} (I^{\hat a})^\gamma_\beta e^\beta   \ , \\
\omega^\mu_\nu = & \hat \omega^\mu_\nu - \e^{\rho_3/3} D_{[\mu} G^b_{\nu]} k^a_b K^a\ ,
\end{aligned}\end{equation}
with the covariant derivatives defined by
\begin{equation}\label{eq:cov_metric_app}\begin{aligned}
 D G^a = & \diff G^a + \tfrac12 t^{ab} \epsilon_{bcd} G^c \wedge G^d \ , \\
 D\rho_3 = & \diff \rho_3 - \epsilon_{abc} t^{bc} G^a\ , \\
 D \rho_4 = & \diff \rho_4 + \epsilon_{abc} t^{bc} G^a  \ , \\
 D k^a_b = & \diff k^a_b - (k^a_{c} \epsilon_{bde} - \tfrac13 k^a_{b} \epsilon_{cde})  t^{cd} G^e \ , \\
 D \zeta^{\hat a}_I = & \diff \zeta^{\hat a}_I - \zeta^{\hat a}_K T^K_{bJ} P^J_I G^b \ .
\end{aligned}\end{equation}
Note that the scalar $\rho_3 + \rho_4$ is ungauged.

In the next section, we compute the Ricci curvature from the Levi-Civita connection. For that we will also need the differential identities
\begin{equation}\label{eq:diff_omega} \begin{aligned}
\diff \omega^I_{\alpha \beta} (\hat e_\mu) = & \tfrac12 G^a_\mu \epsilon_{abc} t^{bc} \omega^I_{\alpha \beta} - T^I_{aJ} G^a_\mu \omega^J_{\alpha \beta}
- \tfrac12 D_\mu \rho_4 \omega^I_{\alpha \beta}
+ \e^{-\rho_4/2} D_\mu \zeta^{\hat a \, I} (I^{\hat a})^\alpha_\beta \\ &
- \zeta^{\hat a \, I}  D_\mu \zeta^{\hat a}_J \omega^J_{\alpha \beta}
+ \e^{-\rho_4/2} \zeta^{\hat a\, I} \zeta^{\hat a}_J T^J_{aK} G^a_\mu \zeta^{\hat b\, K} (I^{\hat b})^\alpha_\beta  \ , \\
\diff \omega^I_{\alpha \beta} (\hat K_a) = & \e^{-\rho_3/3} (k^{-1})^b_a P^I_K T^K_{bL} P^L_J \omega^J_{\alpha \beta}
- \e^{\rho_3/3} k^a_b t^b_J  P^J_K \omega^{I}_{\gamma [\alpha} \omega^{K}_{\beta] \gamma} \ , \\
\nabla^{\theta}_\gamma \omega^I_{\alpha \beta} = 0 \ .
\end{aligned}\end{equation}
that can be computed with help of \eqref{eq:MCexplicit}, \eqref{eq:truncation_ansatz} and \eqref{eq:con_app}. Note that $\nabla^{\theta}$ denotes the $SU(2)$ connection.

\subsection{Ricci curvature}
\label{sec:Ricci}

The components of the Ricci curvature are given by
\begin{equation}\begin{aligned}
{\rm Ric}_{\mu \nu} = & R^\lambda_{\mu \lambda\nu }+  R^a_{\mu a\nu } + R^\alpha_{\mu \alpha \nu} \ , \\
{\rm Ric}_{\mu a} = & R^\nu_{\mu \nu a }+  R^b_{\mu b a } + R^\alpha_{\mu \alpha a} \ , \\
{\rm Ric}_{\mu \alpha} = & R^\nu_{\mu \nu \alpha}+  R^a_{\mu a \alpha } + R^\beta_{\mu \beta \alpha } \ , \\
{\rm Ric}_{ab} = & R^\mu_{a\mu b}+  R^c_{acb} + R^\alpha_{a\alpha b}\ , \\
{\rm Ric}_{a\alpha} = & R^\mu_{a \mu \alpha }+  R^b_{a b \alpha } + R^\beta_{a \beta \alpha }   \ , \\
{\rm Ric}_{\alpha \beta} = & R^\mu_{\alpha \mu \beta }+  R^c_{\alpha c \beta } + R^\gamma_{\alpha \gamma \beta } \ .
\end{aligned}\end{equation}

Let us now compute these components one by one. We start with ${\rm Ric}_{\mu \nu} $. We compute
\begin{equation}\begin{aligned}
R^\lambda_{\mu \lambda \nu } = & {\rm \hat {Ric}}_{\mu\nu} - 3 \e^{2\rho_3/3} g_{3 \, ab} \eta^{\lambda \kappa} D_{[\mu} G^a_{\lambda]} D_{[\nu} G^b_{\kappa]} \ , \\
R^a_{\mu a \nu } = & - \nabla_\nu (D_\mu \rho_3) - \tfrac13 (D_\mu \rho_3)(D_\nu \rho_3) - \tfrac12 {\rm tr} [k^{-1} \cdot (D_\mu k) \cdot k^{-1} \cdot (D_\nu k)]\\ & - \tfrac12 {\rm tr} [(D_\mu k)^T \cdot (g_3^{-1}) \cdot (D_\nu k)] + \e^{2\rho_3/3} g_{3 \, ab} \eta^{\lambda \kappa} D_{[\mu} G^a_{\lambda]} D_{[\nu} G^b_{\kappa]} \ , \\
R^\alpha_{\mu \alpha \nu} = & - \nabla_\nu (D_\mu \rho_4) - \tfrac14 (D_\mu \rho_4) (D_\nu \rho_4) +  (D_\mu \zeta^{\hat a}_I) (D_\nu \zeta^{\hat a\, I})  \ .
\end{aligned}\end{equation}
This gives the external Ricci curvature
\begin{equation}\label{eq:Ric_munu}\begin{aligned}
 {\rm Ric}_{\mu \nu} = & {\rm \hat {Ric}}_{\mu\nu} - \nabla_\nu (D_\mu \rho_4) - \tfrac14 (D_\mu \rho_4) (D_\nu \rho_4) - \nabla_\nu (D_\mu \rho_3) - \tfrac13 (D_\mu \rho_3)(D_\nu \rho_3) \\ &
 -\tfrac12  {\rm tr} [k^{-1} \cdot (D_\mu k) \cdot k^{-1} \cdot (D_\nu k)]
 - \tfrac12 {\rm tr} [(D_\mu k)^T \cdot (g_3^{-1}) \cdot (D_\nu k)] \\ &
 +  (D_\mu \zeta^{\hat a}_I)  (D_\nu \zeta^{\hat a \, I})
  - 2 \e^{2\rho_3/3} g_{3 \, ab} \hat g^{\lambda \kappa} D_{[\mu} G^a_{\lambda]} D_{[\nu} G^b_{\kappa]} \ .
\end{aligned}\end{equation}
For ${\rm Ric}_{\mu a} $ we compute
\begin{equation}\begin{aligned}
R^\nu_{\mu \nu a} = &  -  \nabla^\nu(\e^{\rho_3/3} k^a_b  D_{[\nu} G^b_{\mu]})  - \e^{\rho_3/3} D_{[\mu} G^b_{\nu]} ( \tfrac23 k^a_b D^\nu\rho_3 + \tfrac32 D^\nu k^a_b + \tfrac12(k^{-1})^c_a  D^\nu k^c_d k^d_b ) \ , \\
R^b_{\mu b a} = &  - \tfrac23 \e^{\rho_3/3} D_{[\mu} G_{\nu]}^b k^a_b (D^\nu \rho_3) + \tfrac12 \e^{\rho_3/3} D_{[\mu} G^b_{\nu]} (D^\nu k^a_b + (k^{-1})^c_a (D^\nu k^d_c) k^d_b) \\ &
+\tfrac12 \e^{-\rho_3/3}\epsilon_{abc} k^b_d t^{ed} (D_\mu k^c_e + k^f_e (D_\mu k^f_g) (k^{-1})^g_c) \\ &
- \tfrac12 \e^{-\rho_3/3} (k^{-1})_b^c\epsilon_{cde} t^{de} ((k^{-1})^g_a (D_\mu k^b_g) + (k^{-1})^g_b (D_\mu k^a_g) ) \ , \\
R^\alpha_{\mu \alpha a} = & \tfrac14 \e^{-\rho_3/3}(D_\mu \rho_4) (k^{-1})^b_a \epsilon_{bcd} t^{cd} + \e^{-\rho_3/3}(k^{-1})^b_a T^I_{bJ} \zeta^{\hat a}_I (D_\mu \zeta^{\hat a \, J})
- \e^{\rho_3/3} (D^\nu \rho_4) k^a_b D_{[\mu} G^b_{\nu]}\\ &
-\e^{-\rho_3/3} (k^{-1})^c_b \epsilon_{cde} t^{de} (\tfrac13 (D_\mu \rho_3) \delta^a_b + \tfrac12 ((k^{-1})_b^d  D_\mu k^{a}_d +(k^{-1})_a^d D_\mu k^{b}_d )) \ ,
\end{aligned}\end{equation}
and therefore find
\begin{equation}\begin{aligned}
 {\rm Ric}_{\mu a}  = &    - \e^{\rho_3/3} D_{[\mu} G^b_{\nu]} ( \tfrac43 k^a_b D^\nu\rho_3  + k^a_b  D^\nu \rho_4 + 2 D^\nu k^a_b + (k^{-1})^c_a  D^\nu k^d_c k^d_b ) \\ &
+ \e^{-\rho_3/3} (k^{-1})^b_a T^I_{bJ} \zeta^{\hat a}_I (D_\mu \zeta^{\hat a \, J})
-\tfrac12 \e^{-\rho_3/3}\epsilon_{abc} k^b_d t^{ed} (D_\mu k^c_e + k^f_e (D_\mu k^f_g) (k^{-1})^g_c) \\ &
 -  \nabla^\nu(\e^{\rho_3/3} k^a_b  D_{[\nu} G^b_{\mu]})
+\e^{-\rho_3/3}(k^{-1})^b_a \epsilon_{bcd} t^{cd} (\tfrac14 D_\mu \rho_4 - \tfrac13 D_\mu \rho_3 )   \ .
\end{aligned}\end{equation}
Next we find
\begin{equation}\begin{aligned}
R^\nu_{\mu \nu \alpha} = & 0 \ , \\
R^a_{\mu a \alpha } = & 0  \ , \\
R^\beta_{\mu \beta \alpha } = &  0\ ,
\end{aligned}\end{equation}
where we used in the last equation that $\diff \omega^I = \nabla( \tfrac12 \omega^I_{\beta \alpha} e^\beta) \wedge e^\alpha$. Therefore we find
\begin{equation} \label{eq:Ricmualpha}
{\rm Ric}_{\mu \alpha} = 0 \ .
\end{equation}
Let us now compute the internal component ${\rm Ric}_{ab}$. We find
\begin{equation}\begin{aligned}
R^\mu_{a\mu b} = & -\tfrac13 \nabla^\mu D_\mu \rho_3 \delta_{ab} - \tfrac12 \nabla^\mu ((k^{-1})^c_b D_\mu k_c^a + (k^{-1})^c_a D_\mu k_c^b) - \tfrac19 (D^\mu \rho_3)(D_\mu \rho_3) \delta_{ab}  \\ &
+ \e^{2\rho_3/3} D_{[\mu} G_{\nu]}^c k_c^a D^\mu G^{d\, \nu} k_d^b - \tfrac13 (D^\mu \rho_3) ((k^{-1})^c_b D_\mu k_c^a + (k^{-1})^c_a D_\mu k_c^b) \\ &
+ \tfrac14 (D_\mu k^a_c) g_3^{cd} (D^\mu k^b_d) - \tfrac34 (k^{-1})^c_a (D_\mu k^e_c)  (D^\mu k^e_d) (k^{-1})^d_b \\ &
- \tfrac12 [k^{-1} \cdot (D^\mu k)\cdot k^{-1} \cdot D_\mu k]_{(ab)}\ , \\
R^c_{a c b} = & - \e^{-2 \rho_3/3} (\epsilon_{acd} \epsilon_{bef} (k \cdot t \cdot k^T)^{de} (k \cdot t \cdot k^T )^{fc} + 2(k \cdot t_A\cdot g_3\cdot t \cdot k^T )^{ab} \\ &
\quad  -2 |t| (k \cdot t \cdot k^T)^{(ab)} + \tfrac12 \delta^{ab} |t|^2 )
 - \tfrac16  (D^\mu \rho_3)((k^{-1})^c_b D_\mu k_c^a + (k^{-1})^c_a D_\mu k_c^b) \\ &
+ \tfrac14 ((k^{-1})^d_c D_\mu k_d^a + (k^{-1})^d_a D_\mu k_d^c)((k^{-1})^e_b D_\mu k_e^c + (k^{-1})^e_c D_\mu k_e^b) \\ &
 - \tfrac29 (D^\mu \rho_3)(D_\mu \rho_3)\delta^{ab} \ , \\
R^\alpha_{a \alpha b} = &  \e^{2\rho_3/3} \e^{-\rho_4} k^a_c t^c_I H^{IJ} t^d_J k^b_d
 - \tfrac13 \e^{- 2\rho_3/3} (k^{-1})^c_a  (k^{-1})^d_b \zeta^{\hat a}_I \zeta^{\hat a}_J T^I_{cK} P^{KL}  T^J_{dL} \\ &
 - \tfrac14 \e^{- 2\rho_3/3}\epsilon_{acd}(k \cdot t \cdot k^T)^{cd} \epsilon_{bef} (k \cdot t \cdot k^T)^{ef}
+2  \e^{- 2\rho_3/3} (k \cdot t_A \cdot g_{3} \cdot t \cdot k^T)^{ab} \\ &
 - \tfrac12 D^\mu \rho_4 [((k^{-1})^c_b D_\mu k_c^a + (k^{-1})^c_a D_\mu k_c^b) + \tfrac23 D_\mu \rho_3 \delta^{ab}]\ ,
\end{aligned}\end{equation}
where we defined $g_{3\,ab} = k^c_a k^c_b$, $|t| = g_{3\,ab} t^{ab}$ and $t_A = \tfrac12 (t-t^T)$. Then we get
\begin{equation}\label{eq:Ric_ab}\begin{aligned}
{\rm Ric}_{ab} = & -\tfrac13 \nabla^\mu D_\mu \rho_3 \delta_{ab} - \tfrac12 \nabla^\mu ((k^{-1})^c_b D_\mu k_c^a + (k^{-1})^c_a D_\mu k_c^b)   \\ &
+ \e^{2\rho_3/3} D_{[\mu} G_{\nu]}^c k_c^a D^\mu G^{d\, \nu} k_d^b
+ \tfrac12 (D_\mu k^a_c) g_3^{cd} (D^\mu k^b_d) - \tfrac12 (k^{-1})^c_a (D_\mu k^e_c)  (D^\mu k^e_d) (k^{-1})^d_b \\ &
- (D^\mu \rho_3 + D^\mu \rho_4)[\tfrac12 ((k^{-1})^c_b D_\mu k_c^a + (k^{-1})^c_a D_\mu k_c^b) + \tfrac13 D_\mu \rho_3 \delta^{ab}] \\ &
- \e^{-2 \rho_3/3} (\epsilon_{acd} \epsilon_{bef} (k \cdot t \cdot k^T)^{de} (k \cdot t \cdot k^T )^{fc} -2 |t| (k \cdot t \cdot k^T)^{(ab)} + \tfrac12 \delta^{ab} |t|^2 ) \\ &
 - \tfrac14 \e^{- 2\rho_3/3} \epsilon_{acd}(k \cdot t \cdot k^T)^{cd} \epsilon_{bef} (k \cdot t \cdot k^T)^{ef}
 + \e^{2\rho_3/3} \e^{-\rho_4} k^a_c t^c_I H^{IJ} t^d_J k^b_d  \\ &
 - \tfrac13 \e^{- 2\rho_3/3} (k^{-1})^c_a  (k^{-1})^d_b \zeta^{\hat a}_I \zeta^{\hat a}_J T^I_{cK} P^{KL} T^J_{dL}  \ .
\end{aligned}\end{equation}

Let us now compute the internal component ${\rm Ric}_{a\alpha}$. We find
\begin{equation}\begin{aligned}
R^\mu_{a\mu \alpha} = & 0 \ , \\
R^b_{a b \alpha} = & 0  \ , \\
R^\beta_{a \beta\alpha } = & \nabla^{\theta} \tau^\beta_a (\hat e_\beta, \hat e_\alpha) = 0 \ ,
\end{aligned}\end{equation}
so that
\begin{equation}\label{eq:Ricaalpha}
{\rm Ric}_{a\alpha} =  0 \ .
\end{equation}

Let us now compute the internal component ${\rm Ric}_{\alpha\beta}$. We find
\begin{equation}\begin{aligned}
R^\mu_{ \alpha\mu \beta} = & - \tfrac14 \nabla^{\mu} (D_\mu \rho_4) \delta_{\alpha \beta} + \tfrac12 \nabla^\mu (\e^{\rho_4/2} D_\mu \zeta^{\hat a}_I) P^I_J \omega^J_{\alpha \gamma} (I^{\hat a})^\gamma_\beta \\ &
- \tfrac{1}{16} D_\mu \rho_4 D^\mu \rho_4 \delta_{\alpha \beta}
 - \tfrac14 D^\mu \zeta^{\hat a \, I} D_\mu \zeta^{\hat a}_I \delta_{\alpha \beta}
\\ &
+ \tfrac14 \e^{\rho_4/2} D^\mu \zeta^{\hat a}_I G^a_\mu \epsilon_{abc} t^{bc} \omega^I_{\alpha \gamma} (I^{\hat b})^\gamma_\beta
- \tfrac12 \e^{\rho_4/2} D^\mu \zeta^{\hat a}_I T^I_{aJ} G^a_\mu P^J_K \omega^K_{\alpha \gamma} (I^{\hat b})^\gamma_\beta\\ &
+ \tfrac14 \e^{\rho_4/2} D^\mu \rho_4 D_\mu \zeta^{\hat a}_I \omega^I_{\alpha \gamma} (I^{\hat a})^\gamma_\beta
+ \tfrac12 \e^{\rho_4/2}   D^\mu \zeta^{\hat a}_I T^J_{aK} G^a_\mu \zeta^{\hat a}_J \zeta^{\hat b \, K} \omega^I_{\alpha \gamma} (I^{\hat b})^\gamma_\beta
\ ,
\\
R^a_{\alpha a \beta} = & - \tfrac12 D^\mu \rho_3 (\tfrac12 D_\mu \rho_4 \delta_{\alpha \beta}- \e^{\rho_4/2} D_\mu \zeta^{\hat a}_I \omega^I_{\alpha \gamma} (I^{\hat a})^\gamma_\beta ) \\ &
+ \e^{\rho_3/3} k^a_b t^b_I \omega^I_{\alpha \gamma} (I^{\hat a})^\gamma_\beta \Psi^{\hat a} (\hat K_a)
-\tau^a_\alpha (\hat e_\gamma)\tau^a_\gamma (\hat e_\beta) - \e^{-\rho_3/3} (k^{-1})^b_a \epsilon_{bcd} t^{cd} \tau^a_\alpha (\hat e_\beta) \\ &
+\tfrac12 \e^{-2\rho_3/3} \e^{\rho_4/2}  g_{3}^{-1\, ab} \zeta^{\hat a}_I T^I_{aK} P^K_L T^L_{bM} P^M_J \omega^{J}_{\alpha \gamma} (I^{\hat a})^\gamma_\beta \\ &
-\tfrac12 \e^{\rho_4/2} t^a_I P^I_J \zeta^{\hat a}_K T^K_{aL} P^L_M \omega^{[J}_{\alpha \rho} \omega^{M]}_{\rho \gamma} (I^{\hat a})^\gamma_\beta
 \ ,
\\
R^\gamma_{\alpha \gamma \beta } = & {\rm Ric}^\theta_{\alpha \beta} - \tfrac{3}{16} D^\mu \rho_4 D_\mu \rho_4 \delta_{\alpha \beta}
+ \tfrac14 D^\mu \zeta^{\hat a\, I} D_\mu \zeta^{\hat a}_I \delta_{\alpha \beta}
+ \tfrac14 \e^{\rho_4/2} D^\mu \rho_4 D_\mu \zeta^{\hat a}_I \omega^I_{\alpha \gamma} (I^{\hat a})^\gamma_\beta\\ &
- \e^{\rho_3/3} k^a_b t^b_I \omega^I_{\alpha \gamma} (I^{\hat a})^\gamma_\beta \Psi^{\hat a} (\hat K_a) +\tau^a_\alpha (\hat e_\gamma)\tau^a_\gamma (\hat e_\beta) + \e^{-\rho_3/3} (k^{-1})^b_a \epsilon_{bcd} t^{cd} \tau^a_\alpha (\hat e_\beta) \ ,
\end{aligned}\end{equation}
where ${\rm Ric}^\theta_{\alpha \beta}$ is the Ricci curvature of the $SU(2)$ connection $\theta$. We can compute ${\rm Ric}^\theta_{\alpha \beta}$ by taking the $SU(2)$-covariant derivative of the torsion of the connection $\theta$ \cite{KashaniPoor:2013en}
\begin{equation}
{\rm Ric}^\theta_{\alpha \beta} = \tfrac16 (\diff T^\alpha + \theta^\alpha_\lambda \wedge T^\lambda) (\hat e_\gamma, \hat e_\delta, \hat e_\rho) \epsilon^{\gamma \delta \rho \beta} \ .
\end{equation}
From \eqref{eq:MCexplicit} we find that the component $T^\alpha$ of the torsion torsion tensor of $\theta$ is
\begin{equation}\begin{aligned}
T^\alpha = & \diff e^\alpha + \theta^\alpha_\beta \wedge e^\beta \\
= & - \psi^{\hat a} \wedge (I^{\hat a})^\alpha_\beta e^\beta + K^a \wedge \tau^\alpha_a + \e^\mu \wedge \lambda^\alpha_\mu \ .
\end{aligned}\end{equation}
From this we find for the $SU(2)$ Ricci curvature
\begin{equation}\begin{aligned}
{\rm Ric}^\theta_{\alpha \beta} = & - \tfrac12 \e^{2\rho_3/3} \e^{-\rho_4}  g_{3\, ab} t^a_I H^{IJ} t^b_J  \delta_{\alpha \beta}
- \tfrac12 \e^{-\rho_4/2} \epsilon^{\hat a \hat b \hat c} \zeta^{\hat a}_I \zeta^{\hat b}_J \zeta^{\hat c \, K} t^a_I  T^J_{aK} \delta_{\alpha \beta} \\ &
- \tfrac12 \e^{2\rho_3/3} \e^{-\rho_4/2}  g_{3\, ab} \zeta^{\hat a\, I}  t^a_I  t^b_J P^J_K \omega^K_{\alpha \gamma} (I^{\hat a})^\gamma_\beta
+ \tfrac12 \e^{\rho_4/2} t^a_I P^I_J \zeta^{\hat a}_K T^K_{aL} P^L_M \omega^{[J}_{\alpha \rho} \omega^{M]}_{\rho \gamma} (I^{\hat a})^\gamma_\beta  \\ &
 - \tfrac12 \epsilon^{\hat a \hat b \hat c} \zeta^{\hat a\, I} \zeta^{\hat b}_J  t^a_I  T^J_{aK}P^K_L \omega^L_{\alpha \gamma} (I^{\hat c})^\gamma_\beta  \ .
\end{aligned}\end{equation}
Now we are able to determine ${\rm Ric}_{\alpha \beta}$. We find
\begin{equation}\label{eq:Ric_alphabeta}\begin{aligned}
{\rm Ric}_{\alpha \beta} = &  - \tfrac14 \nabla^{\mu} (D_\mu \rho_4) \delta_{\alpha \beta}  - \tfrac14 D_\mu \rho_4 D^\mu \rho_4 \delta_{\alpha \beta}
- \tfrac14 D^\mu \rho_3 D_\mu \rho_4 \delta_{\alpha \beta} \\ &
- \tfrac12 \e^{2\rho_3/3} \e^{-\rho_4}  g_{3\, ab} t^a_I H^{IJ} t^b_J  \delta_{\alpha \beta}
- \tfrac12 \e^{-\rho_4/2} \epsilon^{\hat a \hat b \hat c} \zeta^{\hat a}_I \zeta^{\hat b}_J \zeta^{\hat c \, K} t^a_I  T^J_{aK} \delta_{\alpha \beta} \\ &
+ \tfrac12 \nabla^\mu (\e^{\rho_4/2} D_\mu \zeta^{\hat a}_I) \omega^I_{\alpha \gamma} (I^{\hat a})^\gamma_\beta + \tfrac12 \e^{\rho_4/2} D^\mu (\rho_4+\rho_3) D_\mu \zeta^{\hat a}_I \omega^I_{\alpha \gamma} (I^{\hat a})^\gamma_\beta\\ &
+ \tfrac14 \e^{\rho_4/2} D^\mu \zeta^{\hat a}_I G^a_\mu \epsilon_{abc} t^{bc} \omega^I_{\alpha \gamma} (I^{\hat b})^\gamma_\beta
- \tfrac12 \e^{\rho_4/2} D^\mu \zeta^{\hat a}_I T^I_{aJ} G^a_\mu P^J_K \omega^K_{\alpha \gamma} (I^{\hat b})^\gamma_\beta\\ &
+ \tfrac12 \e^{\rho_4/2}   D^\mu \zeta^{\hat a}_I T^J_{aK} G^a_\mu \zeta^{\hat a}_J \zeta^{\hat b \, K} \omega^I_{\alpha \gamma} (I^{\hat b})^\gamma_\beta
- \tfrac12 \e^{2\rho_3/3} \e^{-\rho_4/2}  g_{3\, ab} \zeta^{\hat a\, I}  t^a_I  t^b_J P^J_K \omega^K_{\alpha \gamma} (I^{\hat a})^\gamma_\beta\\ &
- \tfrac12 \epsilon^{\hat a \hat b \hat c} \zeta^{\hat b}_I \zeta^{\hat c \, J}  T^I_{a J} t^a_K \omega^K_{\alpha \gamma}(I^{\hat a})^\gamma_\beta
+\tfrac12 \e^{-2\rho_3/3} \e^{\rho_4/2}  g_{3}^{-1\, ab} \zeta^{\hat a}_I T^I_{aK} P^K_L T^L_{bM} P^M_J \omega^{J}_{\alpha \gamma} (I^{\hat a})^\gamma_\beta   \ .
\end{aligned}\end{equation}

From \eqref{eq:Ric_munu}, \eqref{eq:Ric_ab} and \eqref{eq:Ric_alphabeta} we can compute the Ricci scalar. It reads
\begin{equation} \label{eq:Ricscalar}\begin{aligned}
r_{11} = & r_4 - 2 \nabla^\mu (D_\mu \rho_4) - 2 \nabla^\mu (D_\mu \rho_3)
\\ &
 - \tfrac54 (D_\mu \rho_4) (D^\mu \rho_4) - \tfrac43 (D_\mu \rho_3)(D^\mu \rho_3) - 2 D^\mu \rho_4 D_\mu \rho_3
  \\ &
 -\tfrac12  {\rm tr} [k^{-1} \cdot (D_\mu k) \cdot k^{-1} \cdot (D^\mu k)]
 - \tfrac12 {\rm tr} [(D_\mu k)^T \cdot (g_3^{-1}) \cdot (D^\mu k)] \\ &
 + (D_\mu \zeta^{\hat a}_I) (D^\mu \zeta^{\hat a \, I})
  - \e^{2\rho_3/3} g_{3 \, ab}  D_{[\mu} G^a_{\nu]} D^{\mu} G^{b \,\nu}  \\ &
 + \tfrac32  \e^{-2 \rho_3/3} (g_{3\, ab}t^{ab})^2
  - \e^{-2 \rho_3/3} g_{3\, ab} g_{3\, cd} t^{(ac)} t^{(bd)}    - \tfrac{3}{4} \e^{-2 \rho_3/3}  g^{ab}_{3} \epsilon_{acd} \epsilon_{bef} t^{cd} t^{ef} \\ &
 - \e^{2\rho_3/3} \e^{-\rho_4} g_{3\, ab} t^a_I H^{IJ} t^b_J
 - \tfrac13 \e^{- 2\rho_3/3} g_3^{ab} \zeta^{\hat a}_I \zeta^{\hat a}_J T^I_{aK} P^{KL} T^J_{bL}  \\ &
- 2 \e^{-\rho_4/2} \epsilon^{\hat a \hat b \hat c} \zeta^{\hat a \, I} \zeta^{\hat b}_J \zeta^{\hat c \, K} t^a_I  T^J_{aK}  \ .
\end{aligned}\end{equation}

\subsection{Ricci curvature in the Einstein frame}\label{app:RicEinstein}
In order to define the four-dimensional theory in the Einstein frame, we have to perform the Weyl rescaling \eqref{eq:Weyl}. As the scalar fields only depend on four-dimensional spacetime, this mostly affects the Ricci curvature component ${\rm Ric}_{\mu\nu}$, given in \eqref{eq:Ric_munu}. It reads in the four-dimensional Einstein frame
\begin{equation}\begin{aligned}
 {\rm Ric}_{\mu \nu} = & {\rm \tilde {Ric}}_{\mu\nu}  - \tfrac34 (D_\mu \phi) (D_\nu \phi)  - \tfrac12 (D_\mu \rho_3)(D_\nu \rho_3) + \tfrac12 \tilde g_{\mu\nu} \nabla^\rho D_\rho \phi + \tfrac16 \tilde g_{\mu\nu} \nabla^\rho D_\rho \rho_3  \\ &
 -\tfrac12  {\rm tr} [k^{-1} \cdot (D_\mu k) \cdot k^{-1} \cdot (D_\nu k)]
 - \tfrac12 {\rm tr} [(D_\mu k)^T \cdot (g_3^{-1}) \cdot (D_\nu k)] \\ &
 +  (D_\mu \zeta^{\hat a}_I)  (D_\nu \zeta^{\hat a \, I})
  - 2 \e^{2\rho_3/3} g_{3 \, ab} \hat g^{\lambda \kappa} D_{[\mu} G^a_{\lambda]} D_{[\nu} G^b_{\kappa]} \ .
\end{aligned}\end{equation}
The only non-trivial off-diagonal component of the Ricci curvature is
\begin{equation}\begin{aligned}
 {\rm Ric}_{\mu a}  = &     -  \e^{- \rho_3/3}(k^{-1})^b_a \nabla^\nu( \e^{\phi + \rho_3} g_{3\,bc}  D_{[\nu} G^c_{\mu]}) \\ &
+ \e^{-\rho_3/3} (k^{-1})^b_a T^I_{bJ} \zeta^{\hat a}_I (D_\mu \zeta^{\hat a \, J})
-\tfrac12 \e^{-\rho_3/3}\epsilon_{abc} k^b_d t^{ed} (D_\mu k^c_e + k^f_e (D_\mu k^f_g) (k^{-1})^g_c) \\ &
+\e^{-\rho_3/3}(k^{-1})^b_a \epsilon_{bcd} t^{cd} ( \tfrac14 D_\mu \phi -\tfrac12 D_\mu \rho_3)   \ ,
\end{aligned}\end{equation}
which, as we discuss in Section \ref{sec:cons_trunc}, corresponds to the equation of motion for $G^a$.
For the other components of the Ricci curvature we find
\begin{equation}\begin{aligned}
{\rm Ric}_{ab} = & -\tfrac13 \e^{\phi + \rho_3/3} \nabla^\mu D_\mu \rho_3 \delta_{ab} - \tfrac12 \e^{\phi + \rho_3/3} \nabla^\mu ((k^{-1})^c_b D_\mu k_c^a + (k^{-1})^c_a D_\mu k_c^b)   \\ &
+ \e^{2\phi + 4\rho_3/3} D_{[\mu} G_{\nu]}^c k_c^a D^\mu G^{d\, \nu} k_d^b \\ &
+ \tfrac12 \e^{\phi + \rho_3/3}(D_\mu k^a_c) g_3^{cd} (D^\mu k^b_d) - \tfrac12 \e^{\phi + \rho_3/3}(k^{-1})^c_a (D_\mu k^e_c)  (D^\mu k^e_d) (k^{-1})^d_b \\ &
- \e^{- 2\rho_3/3} (\epsilon_{acd} \epsilon_{bef} (k \cdot t \cdot k^T)^{de} (k \cdot t \cdot k^T )^{fc} -2 |t| (k \cdot t \cdot k^T)^{(ab)} + \tfrac12 \delta^{ab} |t|^2 ) \\ &
 - \tfrac14 \e^{- 2\rho_3/3} \epsilon_{acd}(k \cdot t \cdot k^T)^{cd} \epsilon_{bef} (k \cdot t \cdot k^T)^{ef}
 + \e^{2\rho_3/3} \e^{-\rho_4} k^a_c t^c_I H^{IJ} t^d_J k^b_d  \\ &
 - \tfrac13 \e^{- 2\rho_3/3} (k^{-1})^c_a  (k^{-1})^d_b \zeta^{\hat a}_I \zeta^{\hat a}_J T^I_{cK} P^{KL} T^J_{dL}  \ ,\\
{\rm Ric}_{\alpha \beta} = &  - \tfrac14 \e^{\phi +\rho_3/3} \nabla^{\mu} D_\mu (\phi - \tfrac23\rho_3) \delta_{\alpha \beta}
- \tfrac12 \e^{-\phi + 4\rho_3/3}   g_{3\, ab} t^a_I H^{IJ} t^b_J  \delta_{\alpha \beta}\\ &
- \tfrac12 \e^{-\phi/2+\rho_3/3} \epsilon^{\hat a \hat b \hat c} \zeta^{\hat a}_I \zeta^{\hat b}_J \zeta^{\hat c \, K} t^a_I  T^J_{aK} \delta_{\alpha \beta} \\ &
+ \tfrac12 \e^{3\phi/2 }  \nabla^\mu ( D_\mu \zeta^{\hat a}_I) \omega^I_{\alpha \gamma} (I^{\hat a})^\gamma_\beta - \tfrac12 \e^{3\phi/2} D^\mu \zeta^{\hat a}_I T^I_{aJ} G^a_\mu P^J_K \omega^K_{\alpha \gamma} (I^{\hat b})^\gamma_\beta\\ &
+ \tfrac12 \e^{3\phi/2}   D^\mu \zeta^{\hat a}_I T^J_{aK} G^a_\mu \zeta^{\hat a}_J \zeta^{\hat b \, K} \omega^I_{\alpha \gamma} (I^{\hat b})^\gamma_\beta
- \tfrac12 \e^{\rho_3} \e^{-\phi/2}  g_{3\, ab} \zeta^{\hat a\, I}  t^a_I  t^b_J P^J_K \omega^K_{\alpha \gamma} (I^{\hat a})^\gamma_\beta\\ &
- \tfrac12 \epsilon^{\hat a \hat b \hat c} \zeta^{\hat b}_I \zeta^{\hat c \, J}  T^I_{a J} t^a_K \omega^K_{\alpha \gamma}(I^{\hat a})^\gamma_\beta
+\tfrac12 \e^{\phi/2-\rho_3}  g_{3}^{-1\, ab} \zeta^{\hat a}_I T^I_{aK} P^K_L T^L_{bM} P^M_J \omega^{J}_{\alpha \gamma} (I^{\hat a})^\gamma_\beta   \ .
\end{aligned}\end{equation}

Similarly, the eleven-dimensional Ricci scalar \eqref{eq:Ricscalar} transforms to
\begin{equation} \begin{aligned}
g_{\mu\nu} r_{11} = & \tilde g_{\mu\nu} \Big( \tilde r_4 +\nabla^\rho D_\rho \phi + \tfrac13 \nabla^\rho D_\rho \rho_3
 - \tfrac{1}{2} (D_\rho \rho_3) (D^\rho \rho_3)
 - \tfrac{3}{4} (D_\rho \phi)(D^\rho \phi) \\ &
- \tfrac14 {\rm tr} [(D_\rho g_3) \cdot (g_3^{-1}) \cdot (D^\rho g_3) \cdot g_3^{-1}] \\ &
 +  (D_\rho \zeta^{\hat a}_I) (D^\rho \zeta^{\hat a \, I})
  - \e^{\phi + \rho_3} g_{3 \, ab} D_{[\rho} G^a_{\sigma]} D^{\rho} G^{b \,\sigma}  \\ &
 + \tfrac32 \e^{-\phi - \rho_3} (g_{3\, ab}t^{ab})^2
  - \e^{-\phi - \rho_3} g_{3\, ab} g_{3\, cd} t^{(ac)} t^{(bd)}   \\ &
  - \tfrac{3}{4} \e^{-\phi - \rho_3}  g^{ab}_{3} \epsilon_{acd} \epsilon_{bef} t^{cd} t^{ef}
 - \e^{-2\phi + \rho_3} g_{3\, ab} t^a_I H^{IJ} t^b_J\\ &
 - \tfrac13 \e^{-\phi - \rho_3} g_3^{ab} \zeta^{\hat a}_I \zeta^{\hat a}_J T^I_{aK} P^{KL} T^J_{bL}
- 2 \e^{-\tfrac32 \phi } \epsilon^{\hat a \hat b \hat c} \zeta^{\hat a\, I} \zeta^{\hat b}_J \zeta^{\hat c \, K} t^a_I  T^J_{aK} \Big)  \ .
\end{aligned}\end{equation}

\section{Form field computations}

\subsection{Field dualizations}
\label{app:dual}

Here we now explicitly discuss the field dualizations for some of the components of the three-form field $C_3$. The main reason for the dualization is exchanging the two-forms $\hat C_a$ for scalars $\gamma^a$ and dualizing into a standard electric-magnetic duality frame of $N=4$ gauged supergravity, as described in \cite{Schon:2006kz}. As we will see, this requires to exchange the gauge fields $C^a$ for their magnetic duals, which we will denote by $\tilde C_a$. However, to perform the dualization in a consistent way, we also have to introduce further dual auxiliary fields: A magnetic vector field $\tilde C_I$ and the two-forms $\hat C_0$ and $\hat C^a_I$. The magnetic vector field $\tilde C_I$ will appear in the covariant derivative of $\gamma^a$, because $\hat C_a$ appears in the covariant derivative of $C_I$. The new two-forms will appear in the covariant derivative of $\tilde C_a$, since $C^a$ appears in the covariant derivatives of the scalars $c_{aI}$ and $c_0$.

We perform the field dualizations by showing that the set of Bianchi identities and equations of motions are the same, with Bianchi identities swapped for the equations of motion and vice versa. Therefore, let us start by deriving the Bianchi identities from $\diff G_4 = 0 $. This gives for the field strengths $\hat F_a = D \hat C_a - \epsilon_{abc} C^b \wedge DG^c$, $F_I = DC_I + c_{aI} DG^a$ and $F^a = DC^a + c_0 DG^a$ the identities
\begin{equation}\label{eq:Bianchi_original}\begin{aligned}
\diff \hat F_a + \epsilon_{abc} t^{dc} G^b \wedge \hat F_d  + \epsilon_{abc} F^b \wedge DG^c  & = 0 \ , \\
\diff F_I - t^a_I \hat F_a - T^J_{aI} G^a \wedge F_J + \tfrac12 \epsilon_{abc} t^{bc} G^a \wedge F_I - Dc_{aI} \wedge DG^a & = 0 \ , \\
\diff F^a - t^{ba} \hat F_b - Dc_0 \wedge DG^a + t^{ab} \epsilon_{bcd}  G^{c} \wedge F^d - \epsilon_{bcd} t^{cd} G^b \wedge F^a  & = 0 \ .
\end{aligned} \end{equation}
Furthermore, we vary the Lagrangian with respect to the fields $\hat C_a$, $C^a$ and $C_I$ to determine their equations of motion to be
\begin{equation}\label{eq:eomF} \begin{aligned}
0 = & \diff (\e^{2\phi} g_3^{ab} \ast \hat F_b) + t^{ab} \epsilon_{bcd} G^c \wedge (\e^{2\phi} g^{de} \ast \hat F_e)  \\ &
+ \e^{\phi- \rho_3} t^{ab} g_{3\, bc} \ast F^c + \e^{\rho_3} t^a_I H^{IJ} \ast F_J
+ \tfrac12 \epsilon^{abc} \eta^{IJ} Dc_{bI} \wedge Dc_{cJ}   \\ &
+ f_0 DG^a +(g_0 +c_{aI} \eta^{IJ} t^a_J) F^a + (g^a_I + c_0 t^a_I + \epsilon^{abc} T^J_{bI} c_{cJ} + t^{(ab)} c_{bI}) \eta^{IJ} F_J  \ , \\
0 = & (\diff + \epsilon_{bcd} t^{bc} G^d \wedge) (\e^{\phi - \rho_3} g_{3\, ae} \ast F^e) - \epsilon_{abc} t^{db} G^c  \wedge (\e^{\phi - \rho_3} g_{3\, de} \ast F^e) \\ &
- \epsilon_{abc} DG^b \wedge (\e^{2\phi} g_3^{cd} \ast \hat F_d)
+ \epsilon_{abc} t^{bc} (\e^{-2 \rho_3} \ast Dc_0) - f_0 \epsilon_{abc} G^b \wedge DG^c \\ &
- \epsilon_{abc} t^c_I H^{IJ} \e^{-\phi} g^{bd} \ast Dc_{dJ} + (g_0+c_{bI} \eta^{IJ} t^b_J) \hat F^a + \eta^{IJ} D c_{aI} \wedge F_J  \ , \\
0 = & \diff ( \e^{\rho_3} H^J_I \ast F_J) - T^J_{aI} G^a \wedge ( \e^{\rho_3} H^K_J \ast F_K) - \tfrac12 \epsilon_{abc} t^{bc} G^a \wedge ( \e^{\rho_3} H^J_I \ast F_J)\\ &
 - T^J_{aI} H^K_J g_3^{ab} \e^{-\phi} \ast Dc_{bK} - \tfrac12 \epsilon_{abc} t^{bc} H^J_I g^{ad}_3 \e^{-\phi} \ast Dc_{dJ}\\ &
  +(g^a_I + c_0 t^a_I + \epsilon^{abc} T^J_{bI} c_{cJ} + t^{(ab)} c_{bI}) \hat F_a + Dc_{aI} \wedge F^a + Dc_0 \wedge F_I \ .
\end{aligned} \end{equation}
We now want to introduce dual fields with field strengths $\Gamma^a$, $\tilde F_a$ and $\tilde F_I$ in such a way that on-shell the duality relations
\begin{equation}\label{eq:duality}\begin{aligned}
\Gamma^a = & \e^{2\phi} g_3^{ab} \ast \hat F_b \ , \\
\tilde F_a = & \e^{\phi - \rho_3} g_{3\, ab} \ast F^b \ , \\
\tilde F_I = & \e^{\rho_3} H^J_I \ast F_J \ ,
\end{aligned} \end{equation}
should hold. Similarly, we need to introduce two-forms $\hat C_0$ and $\hat C^a_I$ that are related to the scalars $c_0$ and $c_{aI}$ by similar duality relations.
We also need to include seven-form flux $G^{\rm flux}_7$ given in \eqref{eq:G7flux} on the internal space, which will play an important role as soon as we introduce potentials.

From \eqref{eq:eomF} we find the Bianchi identities
\begin{equation}\label{eq:Bianchi_dual} \begin{aligned}
0 = & \diff \Gamma^a + t^{ab} \epsilon_{bcd}\Gamma^c \wedge G^d  + t^{ab} \tilde F_b + t^a_I \eta^{IJ} \tilde F_J
- \tfrac12 \epsilon^{abc} \eta^{IJ} Dc_{bI} \wedge Dc_{cJ}   \\ &
+ f_0 DG^a + (g_0 +c_{aI} \eta^{IJ} t^a_J) F^a + (g^a_I + c_0 t^a_I + \epsilon^{abc} T^J_{bI} c_{cJ} + t^{(ab)} c_{bI}) \eta^{IJ} F_J  \ , \\
0 = & \diff \tilde F_a + \epsilon_{bcd} t^{bc} G^d \wedge\tilde F_a - \epsilon_{abc} t^{db} G^c  \wedge \tilde F_d
+ \epsilon_{abc} \Gamma^b \wedge DG^c + \epsilon_{abc} t^{bc} \hat F_0 \\ &
- \epsilon_{abc} t^c_I \eta^{IJ} \hat F^b_J + (g_0+c_{bI} \eta^{IJ} t^b_J) \hat F^a + \eta^{IJ} D c_{aI}  \wedge F_J - f_0 \epsilon_{abc} G^b \wedge DG^c \ , \\
0 = & \diff \tilde F_I - T^J_{aI} G^a \wedge \tilde F_J - \tfrac12 \epsilon_{abc} t^{bc} G^a \wedge \tilde F_I
 - T^J_{aI} \hat F^a_J - \tfrac12 \epsilon_{abc} t^{bc} \hat F^a_I \\ &
  +(g^a_I + c_0 t^a_I + \epsilon^{abc} T^J_{bI} c_{cJ} + t^{(ab)} c_{bI}) \hat F_a + Dc_{aI} \wedge F^a + Dc_0 \wedge F_I \ .
\end{aligned} \end{equation}
Here $\hat F_0$ and $\hat F^a_I$ denote the field strengths of the two-forms $\hat C_0$ and $\hat C^a_I$, respectively. The Bianchi identities are solved in terms of potentials $\gamma^a$, $\tilde C_a$ and $\tilde C_I$ as well as $\hat C_0$ and $\hat C^a_I$ by
\begin{equation}\label{eq:fieldstrength_dual} \begin{aligned}
 \hat F_0 = & \diff \hat C_0 + \epsilon_{abc} t^{bc} G^a \wedge C_0 - \epsilon_{abc}t^a_I \eta^{IJ} G^b \wedge \hat C^c_J  - \tfrac12 t^a_I \eta^{IJ} C_J \wedge \hat C_a + \tfrac12 \eta^{IJ} C_I \wedge F_J \\ &
 - \tfrac16 f_0 \epsilon_{abc} G^a \wedge G^b \wedge G^c \ , \\
 \hat F^a_I = & \diff \hat C^a_I - T^J_{bI} G^a \wedge \hat C^b_J - \tfrac12 \epsilon_{bcd} t^{cd} G^a \wedge \hat C^b_I -  t^{ba} C_I \wedge \hat C_b  - C^a \wedge F_I - \epsilon^{abc} c_{bI} \hat F_c  \ , \\
 \tilde F_a = & D \tilde C_a - c_{aI} \eta^{IJ} F_J - \epsilon_{abc} \gamma^b DG^c +\tfrac12 c_{aI} \eta^{IJ} c_{bJ} DG^b \ ,\\
 \tilde F_I = & D \tilde C_I - c_0 F_I - c_{aI} F^a + c_0 c_{aI} DG^a \ , \\
 \Gamma^a = & D \gamma^a + \tfrac12 \epsilon^{abc} \eta^{IJ} c_{bI} Dc_{cJ} \ ,
\end{aligned} \end{equation}
where we defined the covariant derivatives
\begin{equation} \begin{aligned}
D \tilde C_a = & \diff \tilde C_a - \epsilon_{abc} t^{bc} \hat C_0 - \epsilon_{abc} t^b_I \eta^{IJ} \hat C^c_J - g_0 \hat C_a - \tfrac12 T^I_{aK} \eta^{KJ} C_I \wedge C_J \\ &
 + \epsilon_{bcd} t^{bc} G^d \wedge\tilde C_a  - \epsilon_{abc} t^{db} G^c \wedge \tilde C_d +\tfrac12 f_0 \epsilon_{abc} G^b \wedge G^c \ , \\
D\tilde C_I = & \diff \tilde C_I + T^J_{aI} \hat C^a_J + \tfrac12 \epsilon_{abc} t^{bc} \hat C^a_I - g^a_I \hat C_a + \tfrac12 t^a_I \epsilon_{abc} C^b \wedge C^c- g^a_I \epsilon_{abc} G^b \wedge C^c \\ &
+ T^J_{aI} C^a \wedge C_J
+ \tfrac12 \epsilon_{abc}t^{bc} C^a \wedge C_I  - T^J_{aI} G^a \wedge \tilde C_J - \tfrac12 \epsilon_{abc} t^{bc} G^a \wedge \tilde C_I
\ , \\
D \gamma^a = & \diff \gamma^a - (f_0 \delta^a_d + t^{ab} \epsilon_{bcd} (\gamma^c + \tfrac12 \epsilon^{cef} c_{eI} \eta^{IJ} c_{fJ})) G^d - t^{ab} \tilde C_b - t^a_I \eta^{IJ} \tilde C_J - g_0 C^a \\ &
- g^a_I \eta^{IJ} C_J
 + c_{bI} \eta^{IJ} ( t^{[a}_J C^{b]} - g^{[a}_J G^{b]} )
+ \tfrac12 \epsilon^{abc}c_{bI} \eta^{IJ} T^K_{cJ} C_K + \tfrac12 t^{[ab]} c_{bI} \eta^{IJ} C_J   \ .
\end{aligned} \end{equation}
By dualizing the Bianchi identities \eqref{eq:Bianchi_original} for $\hat C_a$ and $C^a$ we find the equations of motion for $\tilde C_a$ and $\gamma^a$. They read
\begin{equation}\label{eq:eom_dual}\begin{aligned}
0 = & \diff  ( \e^{-2\phi} g_{3\,ab}\ast \Gamma^b) + \epsilon_{abc} t^{dc} G^b \wedge ( \e^{-2\phi} g_{3\,de}\ast \Gamma^e) - \epsilon_{abc} DG^b \wedge (\e^{\rho_3-\phi} g^{cd}_{3} \ast\tilde F_d)  \ , \\
0 = &\diff (\e^{\rho_3-\phi} g^{ab}_{3} \ast\tilde F_b) - t^{ba} ( \e^{-2\phi} g_{3\,bc}\ast \Gamma^c) + t^{ab} \epsilon_{bcd}  G^{c} \wedge (\e^{\rho_3-\phi} g^{de}_{3} \ast\tilde F_e) \\ & - \epsilon_{bcd} t^{cd} G^b \wedge (\e^{\rho_3-\phi} g^{de}_{3} \ast\tilde F_e) - Dc_0 \wedge DG^a  \ .
\end{aligned} \end{equation}

Now we are in the position to give the dual Lagrangian.
\begin{equation}\begin{aligned}
S_{\rm dual} = - \tfrac{1}{4 \kappa_4^2} \int \Big( & \e^{\rho_3-\phi} g^{ab}_{3}( D \tilde C_a - c_{aI} \eta^{IK} DC_K - \epsilon_{acd} \gamma^c DG^d -\tfrac12 c_{aI} \eta^{IK} c_{cK} DG^c) \\ &
\qquad \wedge  \ast (D \tilde C_b - c_{bJ} \eta^{JL} DC_L - \epsilon_{bef} \gamma^e DG^f -\tfrac12 c_{bJ} \eta^{JL} c_{eL} DG^e)  \\ &
- \e^{-2\phi } g_{3\, ab} (D \gamma^a +\tfrac12 \epsilon^{acd} c_{cI} \eta^{IJ} Dc_{dJ}) \wedge \ast_4 (D \gamma^b+\tfrac12 \epsilon^{bef} c_{eI} \eta^{IJ} Dc_{fJ})
\\ & - 2 c_0 D \tilde C_a \wedge D G^a + g_0 (2DC^a - t^{ba} \hat C_b) \wedge \hat C_a
\\ & +  2 DC^a \wedge \epsilon_{abc} (t^{bc}\hat C_0 + t^b_I \eta^{IJ} \hat C^c_J) - t^a_I\eta^{IJ} (2D \tilde C_J - g^b_J \hat C_b) \wedge \hat C_a
\\ & + \epsilon_{abc} t^{bc} \eta^{IJ} C^a \wedge C_I \wedge DC_J + t^a_I \eta^{IJ} \epsilon_{abc} C^a \wedge C^b \wedge D C_J
\\ & + 2 \epsilon_{abc} t^a_I \eta^{IJ} \tilde C_J \wedge C^b \wedge DG^c + \tfrac16 f_0 \epsilon_{def} t^{ef} \epsilon_{abc} G^a \wedge G^b \wedge G^c \wedge C^d
\Big) \ .
\end{aligned}\end{equation}
The first two terms are the new kinetic terms that replace the ones of $\hat C_a$ and $C^a$. The next term is a topological term to complete the equations of motion of \eqref{eq:eom_dual}. The remaining topological terms ensure that variation with respect to the auxiliary fields gives the duality relations \eqref{eq:duality} as well as\footnote{Note that these equations only arise if the corresponding charges are non-vanishing. If the charges are vanishing, the duality equation does not arise, but also the auxiliary field does not appear in the Lagrangian any more.}
\begin{equation}\begin{aligned}
\hat F_a = & \e^{-2\phi } g_{3\, ab} \ast (D \gamma^b +\tfrac12 \epsilon^{bcd} c_{cJ} \eta^{JK} Dc_{dK}) \ , \\
\hat F_0 = & \e^{-2\rho_3} \ast Dc_0 \ , \\
\hat F^a_I = & \e^{-\phi} H_I^J g^{ab}_3 \ast Dc_{bJ} \ .
\end{aligned} \end{equation}
Note that $\gamma^a$ cannot have a potential term since it inherits the shift symmetry from its dual tensor $\hat C_a$.

\subsection{Energy-Momentum tensor} \label{app:emtensor}

In this appendix we want to compute the eleven-dimensional energy-momentum tensor
\begin{equation}
T_{AB} = \tfrac{1}{12} ( G_{ACDE} G_B{}^{CDE} - \tfrac14 g_{AB} G_{CDEF} G^{CDEF}) \ ,
\end{equation}
that appears in the Einstein field equations \eqref{eq:eom11d}. For this, we use the form \eqref{eq:G4} and the field dualizations that have been discussed in more detail in Section~\ref{sec:fourformred} and in Appendix~\ref{app:dual}.

First of all note that the components $T_{\mu \alpha}$ and $T_{a \alpha}$ both are identical zero, due to the absence of $SU(2)$ doublet degrees of freedom in our ansatz. This fits nicely together with \eqref{eq:Ricmualpha} and \eqref{eq:Ricaalpha}.
For the remaining components, let us first compute $\tilde T_{AB} = \tfrac{1}{12}  G_{ACDE} G_B{}^{CDE}$.  Inserting \eqref{eq:G4} and keeping in mind the Weyl rescaling \eqref{eq:Weyl} we find
\begin{equation}\begin{aligned}
\tilde T_{\mu \nu} = & \tfrac18 g_{\mu\nu} \e^{-3\phi -\rho_3} ( f_0+ (g_0 + c_{aI} \eta^{IJ} t^a_J)c_0 +  (g^a_{I}+ \tfrac12  \epsilon^{abc} T^J_{bI} c_{cJ} + \tfrac12 t^{(ab)}  c_{bI})\eta^{IK} c_{aK} )^2 \\ &
- \tfrac12 \e^{-2\phi} g_{3\, ab} (D_\mu \gamma^a +\tfrac12 \epsilon^{acd} c_{cI} \eta^{IJ} D_\mu c_{dJ}) (D_\nu \gamma^b  +\tfrac12 \epsilon^{bef} c_{eK} \eta^{KL} D_\nu c_{fL}) \\ &
+  \tfrac12 \e^{-2\phi} g_{\mu\nu} (D_\rho \gamma^a +\tfrac12 \epsilon^{acd} c_{cI} \eta^{IJ} D_\rho c_{dJ}) (D^\rho \gamma^b  +\tfrac12 \epsilon^{bef} c_{eK} \eta^{KL} D^\rho c_{fL})  \\ &
+ \tfrac12 \e^{\rho_3} H^{IJ}  F_{I \, \mu\rho} F_{J\, \nu}{}^{\rho} + \tfrac12 \e^{-\phi + \rho_3} g^{3\, ab} \tilde F_a{}_{\mu\rho} \tilde F_b{}_{\nu}{}^{\rho} \\ &
+ \tfrac12 \e^{-2\rho_3} D_\mu c_0 D_\nu c_0 - \tfrac12 \e^{-\phi} H^{IJ} g_3^{ab} D_\mu c_{aI} D_\nu c_{bJ} \ ,
\end{aligned} \end{equation}
\begin{equation}\begin{aligned}
\tilde T_{\mu a} = & \tfrac12 \e^{-2 \phi -\rho_3/3} (f_0+ (g_0 + c_{eI} \eta^{IJ} t^e_J)c_0 + (g^e_{K}+ \tfrac12  \epsilon^{efg} T^M_{fK} c_{gM} + \tfrac12 t^{(ef)}  c_{fK})\eta^{KL} c_{eL} ) \\ & \qquad k^a_b (D_\mu \gamma^b  +\tfrac12 \epsilon^{bcd} c_{cI} \eta^{IJ} D_\mu c_{dJ}) \\ &
+ \tfrac14 \e^{ 2\rho_3/3-\phi} k^d_a\epsilon_{dbc} g_3^{cg} \tilde F_{g\, \mu\nu} (D^\nu \gamma^b +\tfrac12 \epsilon^{bef} c_{eI} \eta^{IJ} D^\nu c_{fJ}) \\ &
+ \tfrac12 \e^{ 2\rho_3/3} (k^{-1})^b_a H^{IJ} F_{I \,\mu\nu} D^\nu c_{bJ} + \tfrac12 \e^{ - \rho_3/3} \epsilon_{\mu\nu\rho\sigma} (k^{-1})^{b}_a \tilde F_b^{\rho \sigma} D^\nu c_0  \\ &
+ \tfrac12 \e^{- \phi - \rho_3/3}k^b_a\epsilon_{bcd} D_\mu c_{cI} H^{IJ} (g^d_{J}+ c_0 t^d_J + \epsilon^{def} T^K_{eJ} c_{fK} + t^{(de)} c_{eJ})\ ,
\end{aligned} \end{equation}
\begin{equation}\begin{aligned}
\tilde T_{ab} = & \tfrac12 \e^{-\phi +\rho_3/3} g_{3\, ac} (D_\mu \gamma^c +\tfrac12 \epsilon^{cef} c_{eI} \eta^{IJ} D_\mu c_{fJ}) g_{3\, bd} (D^\mu \gamma^d  +\tfrac12 \epsilon^{dgh} c_{gK} \eta^{KL} D^\mu c_{hL})\\ &
 + \tfrac14 \e^{4\rho_3/3} (g_{3\, ab} g_3^{cd} - \delta_{a}^c \delta_{b}^d) \tilde F_c^{\mu \nu} \tilde F_{d\, \mu\nu} \\ &
+ \tfrac16 \e^{\phi -5\rho_3/3} D_\mu c_0 D^\mu c_0 \delta_{ab} + \tfrac12 \e^{\rho_3/3} H^{IJ} D_\mu c_{aI} D^\mu c_{bJ} \\ &
+ \tfrac12 \e^{-\phi -2\rho_3/3} H^{IJ} (\delta_{ab} \delta_{cd} - \delta_{ac} \delta_{bd}) (g^c_{I}+ c_0 t^c_I + \epsilon^{cef} T^K_{eI} c_{fK} + t^{(ce)} c_{eI})\\ & \qquad (g^b_{J}+ c_0 t^b_J + \epsilon^{bgh} T^L_{gJ} c_{hL} + t^{(bg)} c_{gJ}) \ ,
\end{aligned} \end{equation}
\begin{equation}\begin{aligned}
\tilde T_{\alpha \beta} = & \tfrac12 \e^{-2 \phi + 4\rho_3/3} (g_0 + c_{aI} \eta^{IJ} t^a_J)^2
+ \tfrac12 g_{3\, ab} \e^{-\phi -2\rho_3/3}(H^{IJ} \delta_{\alpha \beta} - 2 \e^{\rho_4/2} \zeta^{\hat a (I} P^{J)}_K \omega^K_{\alpha \gamma} (I^{\hat a})^\gamma_\beta) \\ & \qquad
(g^a_{I}+ c_0 t^a_I + \epsilon^{acd} T^K_{cI} c_{dK} + t^{(ab)} c_{bI})(g^b_{J}+ c_0 t^b_J + \epsilon^{bef} T^L_{eJ} c_{fL} + t^{(be)} c_{eJ}) \\ &
+ \tfrac12 \e^{\rho_3/3} g_3^{ab} ( H^{IJ} \delta_{\alpha \beta} - 2 \e^{\rho_4/2} \zeta^{\hat a (I} P^{J)}_K \omega^K_{\alpha \gamma} (I^{\hat a})^\gamma_\beta) D_\mu c_{aI} D^\mu c_{bJ} \\ &
+ \tfrac12 \e^{  \phi +4 \rho_3 /3} ( H^{IJ} \delta_{\alpha \beta} - 2 \e^{\rho_4/2} \zeta^{\hat a (I} P^{J)}_K \omega^K_{\alpha \gamma} (I^{\hat a})^\gamma_\beta) F_I^{\mu\nu} F_{J\, \mu\nu} \ .
\end{aligned} \end{equation}
Here we used \eqref{eq:eomhatF} to replace $\hat F$, the scalar-tensor duality relation \eqref{eq:dual_rel_tensors} to replace the tensor fields $\hat C_a$ by their dual scalars $\gamma^a$ and the electro-magnetic duality relation \eqref{eq:dual_rel_vectors} to replace the magnetic vector fields $C^a$ by their electric counterparts $\tilde C_a$.

\section{$N=4$ gauged supergravity}
\label{sec:N=4gaugedSUGRA}
We review here the basic notation of gauged $N=4$ supergravity in the embedding tensor formalism \cite{deWit:2005ub,Schon:2006kz}.
The theory consists of $n+6$ electric and same number of magnetic vector fields $V^{\alpha M}$, where $\alpha=+,-$ denotes electric and magnetic components and $M=1, \dots , n+6$ labels the vector multiplet index. Here $n$ denotes the number of vector multiplets of the theory. The magnetic vectors have no kinetic term and are therefore auxiliary fields in the theory.
The scalar split into one complex scalar $\tau = \tau_1 + \iu \tau_2 $ in the gravity multiplet that specifies the $Sl(2)/SO(2)$ coset matrix
\begin{equation}
 M_{\alpha \beta} = \frac{1}{\tau_2} \left( \begin{aligned} 1 && \tau_1 \\ \tau_1 && |\tau|^2 \end{aligned} \right) \ .
\end{equation}
The vector multiplet scalar fields combine into an $SO(6,n)/(SO(6)\times SO(n))$ coset matrix $M_{MN}$. The flat metric defining $SO(6,n)$ is denoted by $\eta_{MN}$.
To complete the bosonic field content, there are also auxiliary two-form gauge fields $B^{MN} = B^{[MN]}$ and $B^{\alpha \beta} = B^{(\alpha \beta)}$.

The non-trivial data of $N=4$ gauged supergravity are its charges, which are determined by $\xi_{\alpha M}$ and $f_{\alpha MNP} = f_{\alpha [MNP]}$. Then the following combinations of charges occur regularly
\begin{equation}\begin{aligned}
\Theta_{\alpha MNP} = & f_{\alpha MNP} - \xi_{\alpha [N} \eta_{P]M} \ , \\
\hat f_{\alpha MNP} = & f_{\alpha MNP} - \xi_{\alpha [M} \eta_{P]N}- \tfrac32 \xi_{\alpha N} \eta_{MP} \ .
\end{aligned} \end{equation}
The scalar covariant derivatives are then given by
\begin{equation}\begin{aligned}
D M_{\alpha \beta} = & \diff M_{\alpha \beta} + \xi_{(\alpha| M|} M_{\beta) \gamma} V^{M \gamma} - \xi_{\rho M} \epsilon^{\rho \gamma} \epsilon_{\delta (\alpha} M_{\beta) \gamma} V^{M \delta} \ ,\\
D M_{MN} = & \diff M_{MN} + 2 V^{P \alpha} \Theta_{\alpha P (M}{}^Q M_{N)Q} \ .
\end{aligned}\end{equation}
The field strengths of the vector fields are given by
\begin{equation}
D V^{M\alpha} =  \diff V^{M\alpha} - \tfrac12 \hat f_{\beta NP}{}^M V^{N\beta} \wedge V^{P \alpha} + \tfrac12 \epsilon^{\alpha \beta} \Theta_{\beta} {}^M{}_{NP} B^{NP} + \tfrac12 \xi_{\beta}^M B^{\alpha \beta} \ ,
\end{equation}
with $\epsilon^{+-} = 1$.
The kinetic terms are then given by
\begin{equation}\begin{aligned}
 S_{\rm kin} = \tfrac{1}{2\kappa_{4}^2}  \int   \Big( & (\ast 1) r - \tfrac14 D M_{MN} \wedge \ast D M^{MN} - \tfrac18 D M_{\alpha \beta} \wedge \ast D M^{\alpha \beta} \\ &
 - 2 \Im (\tau) M_{MN} D V^{M+} \wedge \ast D V^{N+} + \Re (\tau) \eta_{MN}D V^{M+} \wedge D V^{N+} \Big) \ ,
\end{aligned}\end{equation}
and the topological terms are given by
\begin{equation}\begin{aligned}
 S_{\rm top} = \tfrac{1}{2\kappa_{4}^2}  \int   \Big( & \xi_{+M} \eta_{NP} V^{M-} \wedge V^{N+} \wedge \diff V^{P+} \\ &
 - (\hat f_{-MNP} + 2 \xi_{-N} \eta_{MP} ) V^{M-} \wedge V^{N+} \wedge \diff V^{P-} \\ &
 -\tfrac14 \hat f_{\alpha MNR} \hat f_{\beta PQ}{}^R V^{M\alpha} \wedge V^{N+} \wedge V^{P\beta}\wedge V^{Q-} \\ &
 + \tfrac{1}{16} \Theta_{+MNP} \Theta_{-}{}^M{}_{QR} B^{NP} \wedge B^{QR} \\ &
 -\tfrac12 (\Theta_{-MNP} B^{NP} + \xi_{\alpha M} B^{+\alpha})\wedge (\diff V^{M-} - \tfrac12 \hat f_{\alpha QR}{}^M V^{Q\alpha} \wedge V^{R-}) \Big)   \ .
\end{aligned}\end{equation}
The scalar potential of the theory is given by
\begin{equation}\begin{aligned}
 S_{\rm pot} = \tfrac{1}{16\kappa_{4}^2}  \int  (\ast 1) \Big( & f_{\alpha MNP} f_{\beta QRS} M^{\alpha \beta} ( \tfrac13 M^{MQ} M^{NR}M^{PS} \\ &
  +(\tfrac23 \eta^{MQ} - M^{MQ}) \eta^{NR} \eta^{PS} ) \\ &
  - \tfrac49 f_{\alpha MNP} f_{\beta QRS} \epsilon^{\alpha \beta} M^{MNPQRS} + 3 \xi^M_\alpha \xi^N_\beta M^{\alpha \beta} M_{MN} \Big)  \ ,
\end{aligned}\end{equation}
where we defined the totally antisymmetric tensor
\begin{equation}
 M_{MNPQRS} = \epsilon^{mnpqrs} \nu^M_m \nu^N_n \nu^P_p \nu^Q_q \nu^R_r \nu^S_s \ ,
\end{equation}
from the $SO(6,n)$ vielbein $\nu^M_m$.
Finally, the embedding tensor components obey a number of quadratic constraints that are necessary in order to ensure locality of the supergravity. These constraints are given by
\begin{equation}\label{eq:quad_constr}\begin{aligned}
\xi^M_\alpha \xi_{\beta M} & = 0 \ , \\
\xi^P_{(\alpha} f_{\beta) PMN} & = 0 \ , \\
3 f_{\alpha R[MN} f_{\beta PQ]}{}^R + 2 \xi_{(\alpha[M} f_{\beta) NPQ] } & = 0 \ , \\
\epsilon^{\alpha \beta} \Big( \xi^P_\alpha f_{\beta PMN} + \xi_{\alpha M} \xi_{\beta N}\Big) & = 0 \ , \\
\epsilon^{\alpha \beta} \Big( f_{\alpha MNR} f_{\beta PQ}{}^R - \xi^R_\alpha f_{\beta R[M[P} \eta_{Q]N]} - \xi_{\alpha [M} f_{\beta |N][PQ]} + \xi_{\alpha [P} f_{\beta|Q][MN]} \Big) & = 0 \ .
\end{aligned}\end{equation}


\begin{thebibliography}{10}


\bibitem{KashaniPoor:2013en}
  A.~-K.~Kashani-Poor, R.~Minasian and H.~Triendl,
  ``Enhanced supersymmetry from vanishing Euler number,''
  JHEP {\bf 1304}, 058 (2013)
  [arXiv:1301.5031 [hep-th]].

\bibitem{ReidEdwards:2008rd}
  R.~A.~Reid-Edwards and B.~Spanjaard,
  ``N=4 Gauged Supergravity from Duality-Twist Compactifications of String Theory,''
  JHEP {\bf 0812} (2008) 052
  [arXiv:0810.4699 [hep-th]].

\bibitem{Triendl:2009ap}
  H.~Triendl and J.~Louis,
  ``Type II compactifications on manifolds with SU(2) x SU(2) structure,''
  JHEP {\bf 0907} (2009) 080
  [arXiv:0904.2993 [hep-th]].

\bibitem{Louis:2009dq}
  J.~Louis, D.~Martinez-Pedrera and A.~Micu,
  ``Heterotic compactifications on SU(2)-structure backgrounds,''
  JHEP {\bf 0909} (2009) 012
  [arXiv:0907.3799 [hep-th]].

\bibitem{Danckaert:2011ju}
  T.~Danckaert, J.~Louis, D.~Martinez-Pedrera, B.~Spanjaard and H.~Triendl,
  ``The N=4 effective action of type IIA supergravity compactified on SU(2)-structure manifolds,''
  JHEP {\bf 1108} (2011) 024
  [arXiv:1104.5174 [hep-th]].

\bibitem{Grimm:2014aha}
  T.~W.~Grimm, A.~Kapfer and S.~Lust,
  ``Partial Supergravity Breaking and the Effective Action of Consistent Truncations,''
  JHEP {\bf 1502}, 093 (2015)
  [arXiv:1409.0867 [hep-th]].


\bibitem{deWit:2005ub}
  B.~de Wit, H.~Samtleben and M.~Trigiante,
  ``Magnetic charges in local field theory,''
  JHEP {\bf 0509} (2005) 016
  [hep-th/0507289].



\bibitem{Schon:2006kz}
  J.~Schon and M.~Weidner,
  ``Gauged N=4 supergravities,''
  JHEP {\bf 0605} (2006) 034
  [arXiv:hep-th/0602024].


\bibitem{ethomas}
E.~ Thomas, ``Vector fields on manifolds,'' Bull. Amer. Math. Soc. 75 (1969), 643-
683.

\bibitem{Friedrich:1997}
  T.~Friedrich, I.~Kath, A.~Moroianu and U.~Semmelmann,
  ``On nearly parallel G2-structures,''
  J.\ Geom.\ Phys.\ {\bf 23}, 259 (1997)
  [hal-00126037].

\bibitem{Kaste:2003zd}
  P.~Kaste, R.~Minasian and A.~Tomasiello,
  ``Supersymmetric M theory compactifications with fluxes on seven-manifolds and G structures,''
  JHEP {\bf 0307}, 004 (2003)
  [hep-th/0303127].


\bibitem{Cassani:2010uw}
  D.~Cassani, G.~Dall'Agata and A.~F.~Faedo,
  ``Type IIB supergravity on squashed Sasaki-Einstein manifolds,''
  JHEP {\bf 1005}, 094 (2010)
  [arXiv:1003.4283 [hep-th]].

\bibitem{Gauntlett:2010vu}
  J.~P.~Gauntlett and O.~Varela,
  ``Universal Kaluza-Klein reductions of type IIB to N=4 supergravity in five dimensions,''
  JHEP {\bf 1006}, 081 (2010)
  [arXiv:1003.5642 [hep-th]].

\bibitem{Cassani:2010na}
  D.~Cassani and A.~F.~Faedo,
  ``A Supersymmetric consistent truncation for conifold solutions,''
  Nucl.\ Phys.\ B {\bf 843}, 455 (2011)
  [arXiv:1008.0883 [hep-th]].

\bibitem{Bena:2010pr}
  I.~Bena, G.~Giecold, M.~Grana, N.~Halmagyi and F.~Orsi,
  ``Supersymmetric Consistent Truncations of IIB on $T^{1,1}$,''
  JHEP {\bf 1104}, 021 (2011)
  [arXiv:1008.0983 [hep-th]].

\bibitem{OColgain:2011ng}
  E.~O Colgain and O.~Varela,
  ``Consistent reductions from D=11 beyond Sasaki-Einstein,''
  Phys.\ Lett.\ B {\bf 703}, 180 (2011)
  [arXiv:1106.4781 [hep-th]].

\bibitem{Cassani:2011fu}
  D.~Cassani and P.~Koerber,
  ``Tri-Sasakian consistent reduction,''
  JHEP {\bf 1201}, 086 (2012)
  [arXiv:1110.5327 [hep-th]].

\bibitem{Cassani:2012pj}
  D.~Cassani, P.~Koerber and O.~Varela,
  ``All homogeneous N=2 M-theory truncations with supersymmetric AdS4 vacua,''
  JHEP {\bf 1211}, 173 (2012)
  [arXiv:1208.1262 [hep-th]].


\bibitem{joyce}
   D.~Joyce, ``Compact Riemannian $7$-manifolds with holonomy $G_2$. I, II,''
J. Differential Geom., 43, 1996, 2.

\bibitem{Papadopoulos:1995da}
  G.~Papadopoulos and P.~K.~Townsend,
  ``Compactification of D = 11 supergravity on spaces of exceptional holonomy,''
  Phys.\ Lett.\ B {\bf 357}, 300 (1995)
  [hep-th/9506150].


\bibitem{Witten:1996md}
  E.~Witten,
  ``On flux quantization in M theory and the effective action,''
  J.\ Geom.\ Phys.\  {\bf 22}, 1 (1997)
  [hep-th/9609122].

\bibitem{Kachru:2004jr}
  S.~Kachru and A.~-K.~Kashani-Poor,
  ``Moduli potentials in type IIa compactifications with RR and NS flux,''
  JHEP {\bf 0503}, 066 (2005)
  [hep-th/0411279].

\bibitem{Louis:2014gxa}
  J.~Louis and H.~Triendl,
  ``Maximally supersymmetric AdS$_{4}$ vacua in N = 4 supergravity,''
  JHEP {\bf 1410}, 007 (2014)
  [arXiv:1406.3363 [hep-th]].


\end{thebibliography}
\end{document}